\documentclass[11pt]{article}
\pdfoutput=1
\usepackage{jcappub}
\usepackage{graphicx,amssymb,afterpage}
\usepackage{siunitx}
\graphicspath{{figs/}}

\newcommand{\fig}[2]{\includegraphics[width=#1\textwidth]{#2}}
\newcommand{\figh}[2]{\includegraphics[height=#1\textheight]{#2}}

\usepackage[colorinlistoftodos]{todonotes}

\begin{document}

\title{An extensive-air-shower-like event registered with the TUS
orbital detector}

\author[a]{B.A.~Khrenov,}
\author[a]{G.K. Garipov,}
\author[a]{M.A. Kaznacheeva,}
\author[a]{P.A.~Klimov,}
\author[a,b]{M.I.~Panasyuk,}
\author[a]{V.L.~Petrov,}
\author[a]{S.A.~Sharakin,}
\author[a]{A.V.~Shirokov,}
\author[a]{I.V.~Yashin,}
\author[a]{M.Yu.~Zotov,}
\author[c]{A.A.~Grinyuk,}
\author[c,d]{V.M.~Grebenyuk,}
\author[c]{M.V.~Lavrova,}
\author[c,d]{L.G.~Tkachev,}
\author[c]{A.V.~Tkachenko,}
\author[e]{O.A.~Saprykin,}
\author[e]{A.A.~Botvinko,}
\author[e]{A.N.~Senkovsky,}
\author[e]{A.E.~Puchkov,}
\author[f]{M.~Bertaina}
\author[f]{and A.~Golzio}

\affiliation[a]{Lomonosov Moscow State University, Skobeltsyn Institute
of Nuclear Physics, GSP-1, Leninskie Gory, Moscow, 119991, Russia}

\affiliation[b]{Physics Department, Lomonosov Moscow State University,
Leninskie Gory, Moscow, 119991, Russia}

\affiliation[c]{Joint Institute for Nuclear Research,
Joliot-Curie, 6, Dubna, 141980, Moscow region, Russia}

\affiliation[d]{Dubna State University,
University str., 19, Bld.1, Dubna, Moscow region, Russia}

\affiliation[e]{Space Regatta Consortium,
ul. Lenina, 4a, Korolev, 141070, Moscow region, Russia}

\affiliation[f]{Universit\`a degli studi di Torino,
Via Pietro Giuria 1, 10125 Turin, Italy}

\emailAdd{pavel.klimov@gmail.com, sharakin@mail.ru,
zotov@eas.sinp.msu.ru}

\abstract{TUS (Tracking Ultraviolet Set-up) is the world's first orbital
detector of ultra-high-energy cosmic rays (UHECRs). It was launched into
orbit on 28th April 2016 as a part of the scientific payload of the
Lomonosov satellite. The main aim of the mission was to test the
technique of measuring the ultraviolet fluorescence and Cherenkov
radiation of extensive air showers generated by primary cosmic rays with
energies above $\sim100$~EeV in the Earth atmosphere from space. During
its operation for 1.5 years, TUS registered almost 80,000 events with a few of
them satisfying conditions anticipated for extensive air showers (EASs)
initiated by UHECRs.  Here we discuss an event registered on 3rd~October
2016.  The event was measured in perfect observation conditions as an
ultraviolet track in the nocturnal atmosphere of the Earth, with the
kinematics and the light curve similar to those expected from an EAS.  A
reconstruction of parameters
of a primary particle
gave the zenith angle around $44^\circ$ but an extreme energy
not compatible with the cosmic ray energy
spectrum obtained with ground-based experiments.
We discuss in details all conditions of registering the event, explain the
reconstruction procedure and its limitations and comment on
possible sources of the signal, both of anthropogenic and astrophysical
origin.
We believe this detection represents a significant milestone in the
space-based observation of UHECRs because it proves the capability of
an orbital telescope to detect light signals with the apparent motion and
light shape similar to what are expected from EASs.
This is important for the on-going development of the future missions
KLYPVE-EUSO and POEMMA, aimed for studying UHECRs from space.  }


\arxivnumber{1907.06028}

\maketitle

\section{Introduction}

Measurements of the energy spectrum, nuclear composition and arrival
directions of ultra-high-energy cosmic rays (UHECRs,
$E\gtrsim50$~EeV\footnote{1~EeV $= 10^{18}$~eV}) are an important part
of the modern astrophysics and particle physics~\cite{Dawson_etal-2017}.  It
was more than 50 years ago that the first cosmic ray particles of so
extreme energies were detected~\cite{1961PhRvL...6..485L} and a cut-off
of the energy spectrum was predicted~\cite{Greisen-1966,ZK-1966}.
However, the nature and origin of UHECRs are still not understood.  To a
great extent, the problem relates to their very low flux.  Suffices to
say the biggest UHECR experiments---the Pierre Auger Observatory and the
Telescope Array---registered less than two dozen events with energies
$E>100$~EeV in 13 and 7 years of operation
respectively~\cite{Auger-spectrum-2017,TA-spectrum-ICRC-2017}.

The primary goal of the TUS project, first announced in
2001~\cite{2001AIPC..566...57K}, was to expand the UHECR experimental
studies to space as suggested by Benson and Linsley in early
1980's~\cite{1980BAAS...12Q.818B,Benson-Linsley-1981}. The main idea is
that fluorescence and Cherenkov ultraviolet (UV) radiation of an
extensive air shower (EAS) generated by an UHECR in the nocturnal
atmosphere of the Earth can be detected from a satellite similar to the way
it is observed from the ground with fluorescence telescopes but with
a much larger exposure, thus considerably increasing the number of
registered events.

The TUS instrument on board the Lomonosov satellite was launched into
orbit from the newly built Vostochny Cosmodrome (Russia) on 28th April
2016.  The satellite had a sun-synchronous orbit with an inclination of
$97.3^\circ$, a period of $\approx94$~min, and a height of about
470--500~km above the sea level~\cite{tus-jcap-2017,Zotov:uhecr2016}.
The detector operated till late 2017 and registered almost 80,000 events
in the mode aimed at studying UHECRs, see below.  The upper limit of the
total exposure is $\sim2000~\text{km}^2~\text{sr~yr}$. The estimate
reduces to $\sim1200\text{--}1400~\text{km}^2~\text{sr~yr}$ after one
takes into account different penalty factors that arise due to a high
background illumination (and thus a higher energy threshold) over
thunderstorm regions, auroral and urban areas, during periods of
full-moon nights etc.\ assuming their weights are the same as for the
JEM-EUSO project~\cite{2019EPJWC.21006006B,JEM-EUSO-exposure}.

Several preliminary UHECR candidates have been selected in the TUS data
earlier~\cite{2017arXiv170605369B,2019EPJWC.21006006B}.
We found six events among them with the shape and kinematics of the
signals that resemble what is expected from an EAS. In five of these
cases, the field of view (FOV) of the hit pixels correlated with location
of airports thus strongly suggesting an anthropogenic origin of the
signals.
In the present work, we focus on an event registered on 3rd October 2016
(TUS161003 for short), which is different from the other five since
there is neither an airport near the FOV of the hit pixels, nor another
obvious man-made object or settlement that could be the source of a
sufficiently strong emission in the UV range.
First, we describe the experiment, then present the event and discuss
its phenomenology, temporal and spatial behavior and conditions of the
observation. Next, we focus on the reconstruction of the arrival
direction and energy of a possible source of the event assuming
it had an UHECR origin.
Finally, we discuss the result including possible anthropogenic
and astrophysical sources of the signal.


\section{The TUS detector}

\subsection{Design of the detector}

TUS is a UV telescope looking downward into the atmosphere in the nadir
direction. It consists of the two main parts: a modular Fresnel
mirror-concentrator and 256 photomultiplier tubes (PMTs) arranged in a
$16\times16$-channel photodetector located in the focal plane of the
mirror.  The overall field of view of the detector is
$9^\circ\times9^\circ$.

The TUS mirror is composed of~7 hexagonal segments made of a carbon
plastic with the total area of $S_\mathrm{mirr} =1.93$~m$^2$. The focal
distance of the mirror equals 1500~mm. The local reflectivity of the
mirror is higher than 85\% in the near-UV range.

The photodetector is built of 16 modules, each consisting of 16
channels.  Each channel (pixel) is a Hamamatsu R1463 PMT with the
quantum efficiency of approximately~20\% in the wavelength band
300--400~nm.  A multi-alcali cathode is covered by a UV glass filter of
the UFS1 type and a reflective light guide with a square entrance of
the 15~mm size located in the focal plane of the mirror.  The angular
resolution (the FOV of one channel) equals 10~mrad, which corresponds to a
spatial domain of about $5~\text{km}\times5~\text{km}$ at the sea level
from a 500~km orbit height. Thus, the full instantaneous area observed
by TUS on the ground is approximately $80~\text{km}\times80~\text{km}$.

The general design of TUS is determined by its main task of registering
fluorescence and backscattered (reflected) Cherenkov radiation of EASs
in the near-UV band with a time resolution of 0.8~$\mu$s in a full
temporal interval of 256 time steps.  The temporal and spatial
resolution of the detector makes it possible to measure a time-dependent
signal from an EAS as a source moving rectilinearly at the speed of
light. The trajectory of such a source is the EAS axis, which in its
turn coincides with the arrival direction of a primary particle given
by the azimuth~$\phi$ and the zenith angle~$\theta$.

The mirror projects this trajectory in a rectilinear track resulting in
a moving EAS image on the focal plane.  At any moment of time, the
signal is distributed over several neighbouring channels due to the
non-ideal focusing of the mirror.  The shape of a spot is determined by
the position of the centre of the image and by the point spread function
(PSF) of the mirror.  The PSF was measured during pre-flight tests of
the mirror for different angles of incidence~\cite{ICRC2013TUS}. A
typical root-mean-square radius of a spot varies from approximately
7--8~mm on the axis to 8--10~mm at the edge of the FOV (at $4.5^\circ$),
where the shape of the spot is asymmetric due to coma aberration.

The effectiveness of the TUS optics was estimated to be of the order
of~0.7 based on tests performed at the stage of manufacturing.  This
means about~70\% of all UV photons approaching the entrance pupil are
focused in the PMTs.

The TUS electronics can operate in four modes intended for registering
various fast optical phenomena at different time scales with various
time sampling. In addition to the EAS mode with the 0.8~$\mu$s temporal
resolution, TUS performed observations with the sampling time of
25.6~$\mu$s, 0.4~ms and 6.6~ms. The latter one, a so called ``Meteor'' mode,
is used in the present work for a relative calibration of PMTs.
Every data record contains 256 waveforms each consisting of ADC codes
registered in 256 consecutive moments of time.

The TUS on-line selection system is provided by a two-level
trigger~\cite{2013JPhCS.409a2105G,TUS-sim-ApP-2017}, which allows
selecting events in terms of both the intensity of the signal (a
threshold trigger) and the specific space-time pattern (an adjacency
trigger).  Data of all 256 channels are recorded in case conditions of
both triggers are satisfied thus forming a TUS event. 

\begin{figure}[!ht]
    \centerline{\figh{.27}{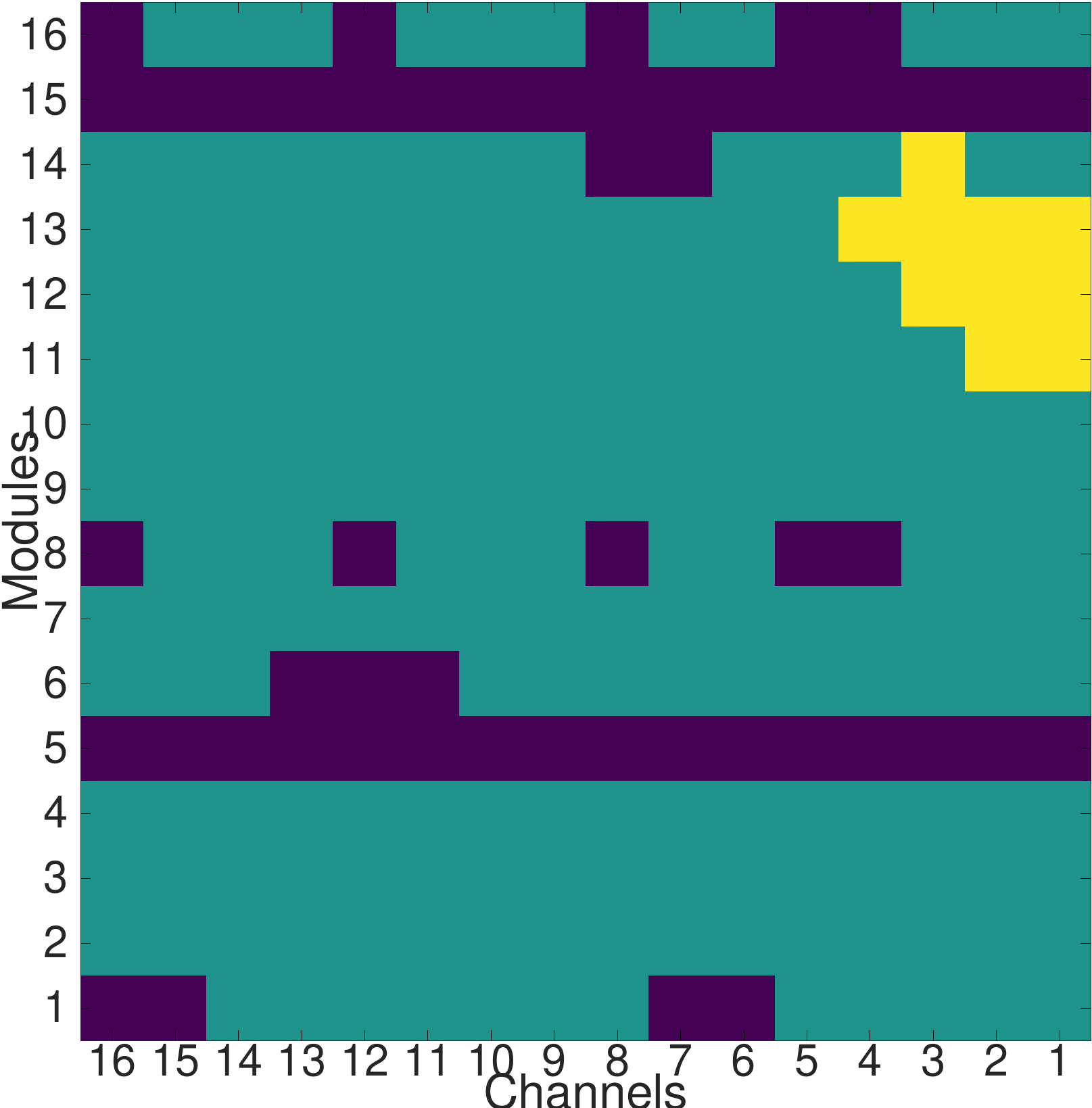}\quad
    \figh{.27}{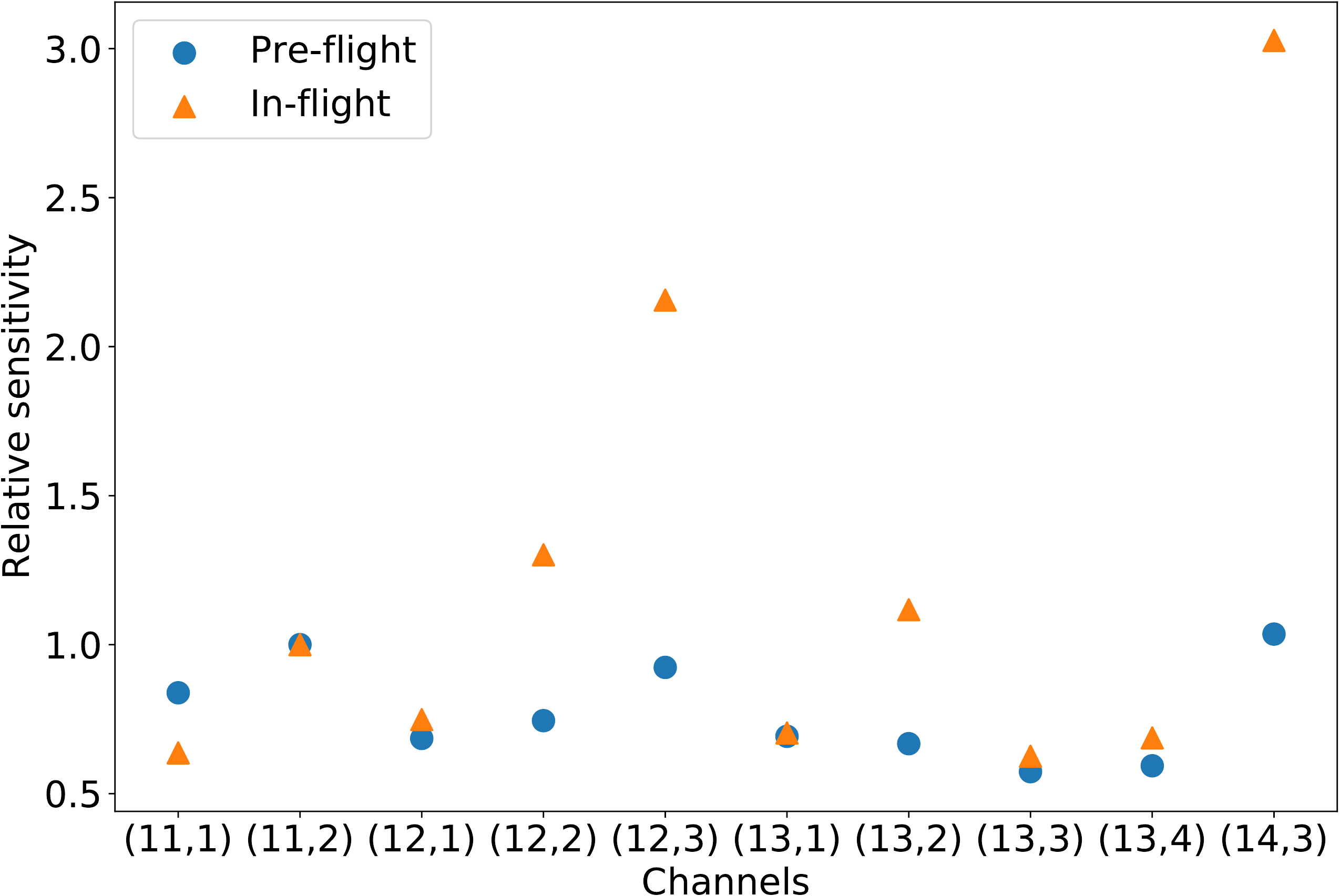}}
    \caption{Left: Map of the channels comprising the focal surface of the
    detector. Dark pixels are dead. The hit channels of the TUS161003 event
    are shown in yellow.
	 Right: relative sensitivity of the hit channels w.r.t.\ channel
	 (11,2) as
    measured during the pre-flight calibration (circles)
    and as estimated with the flight data (triangles).
	 Channels are marked as (module number, channel number).
	 }
    \label{fig:focal_surface}
\end{figure}

The sensitivity of the PMTs was controlled by the high-voltage system,
which was aimed to decrease their sensitivity in conditions of a high
illumination.  Unfortunately, the detector was accidentally operated
at the highest voltage due to a malfunction of the system during the
first few orbits, including both day and nocturnal segments. The reason
of such operation was that the focal surface was partially illuminated
by a very bright Sun light, which caused a shortcut in the high voltage
power supply (HVPS) divider. The HVPS of two modules (number 5 and
number 15) were burnt and these modules did not operate anymore.
In other modules, only PMTs with a high current were damaged but
they had protected other PMTs due to a shortcut because the voltage
was significantly decreased.
Totally, 51 photo detector channels were burnt.
A map of PMTs is shown in the left panel
of Fig.~\ref{fig:focal_surface}. Dark pixels are dead ones,
the other pixels were operating, with the hit channels of the TUS161003
event shown in yellow. One can see the measured event lies in a ``good''
part of the focal surface, with only the starting point of the event
being adjacent to the non-working module.

While 51 PMTs were totally broken by the time of the event measurement,
gains of many other PMTs might have changed so that the validity of the
pre-flight calibration~\cite{SSR} was lost. 
To overcome the problem, a new method of absolute calibration was
developed for estimating the sensitivity of channels of the photodetector.
The method is a combination of the PMT calibration based on an analysis
of the stationary background glow and an estimation of the relative
sensitivity of channels employing events registered in the slow (``Meteor'')
mode.
A change of PMT gains in the hit channels 
as estimated by the method of in-flight calibration versus the
pre-flight values
is demonstrated in the right panel of Fig.~\ref{fig:focal_surface},
see Appendix~\ref{Calibr} for details.

\subsection{Expected detector response to UHECRs}

At any moment of time, UV illumination from an EAS is focused by the optical
system of TUS on a number of adjacent channels (``hit channels'').
A response of the $i$th channel to a signal in 256 time steps of a data
record is represented by 10-bit
ADC codes $A_i(k)$, $k=1,\ldots,256$.  The average value of the
background illumination in each channel (the base level) can be
estimated from a stationary part of the waveform occupying $\sim60$
first time steps of each record.

The digital signal is determined by the emission of an EAS,
transparency of the atmosphere and by the sensitivity of the detector in
a given direction. A ``physical'' signal is represented by an
intensity of the illumination at the entrance pupil~$I_\mathrm{EP}$
expressed as the number of photons per unit time per unit area.  In
particular, by the light curve we mean the dependence
of~$I_\mathrm{EP}$ on time.
A contribution of each $i$th hit channel to $I_\mathrm{EP}$ depends on
its sensitivity~$s_i$:
\begin{equation}\label{eq:Sign}
	I_\mathrm{EP}(k) = \sum_i \frac{A_i(k)}{s_i}.
\end{equation}
Here, the sum is calculated over all hit channels.

In its turn, the sensitivity~$s_i$ of the $i$th channel in the DC-mode
of a PMT can be expressed as a function of the PMT's gain~$G_i$ and the
overall optical efficiency~$\epsilon_i$, which includes the
photo-cathode quantum efficiency, as well as the efficiency of the
mirror and the light guide of the channel:
\begin{equation}\label{eq:s-epsG}
	s_i = a q_\mathrm{e} R 
	  S_\mathrm{mirr} \cos\gamma \cdot\epsilon_i G_i,
\end{equation}
where $a=512$~V$^{-1}$ is the anode voltage-to-ADC code coefficient,
$q_\mathrm{e}$ is the fundamental charge, $R=20$~kOhm is the anode
resistance, and~$\gamma$ is the field angle of the channel in the FOV.

The sensitivity in the nadir direction%
\footnote{The cosine of the maximum field angle $\gamma=4.5^\circ$
equals 0.997, so the angular dependence of the sensitivity can be
neglected.}
is estimated as~0.44~$\mu$s~m$^2$ for the average value of the overall
optical efficiency~0.14, which is the product of the quantum efficiency
equal to~0.2 and the efficiency of the mirror and light guides 
$\approx0.7$, and the PMT gain equal to $10^6$, which is a typical
value for the type of PMTs deployed in the TUS
photodetector~\cite{hamamatsu}.  

Characteristic features of the detector response to a signal from an
EAS can be found with numerical simulations. We utilized the ESAF
(EUSO Simulation and Analysis Framework)~\cite{Berat2010} software to
simulate fluorescence and Cherenkov light from extensive air showers as
they are observed by~TUS.
Detailed simulations of the detector response were performed with the
code developed at University of Turin (Italy)~\cite{2019EPJWC.21006006B}.
Another implementation was developed at Joint Institute for Nuclear
Research (Dubna, Russia) at the R\&D stage~\cite{TUS-sim-ApP-2017}.  Both
codes provide similar outcomes. 

\begin{figure}[!t]
	\fig{.44}{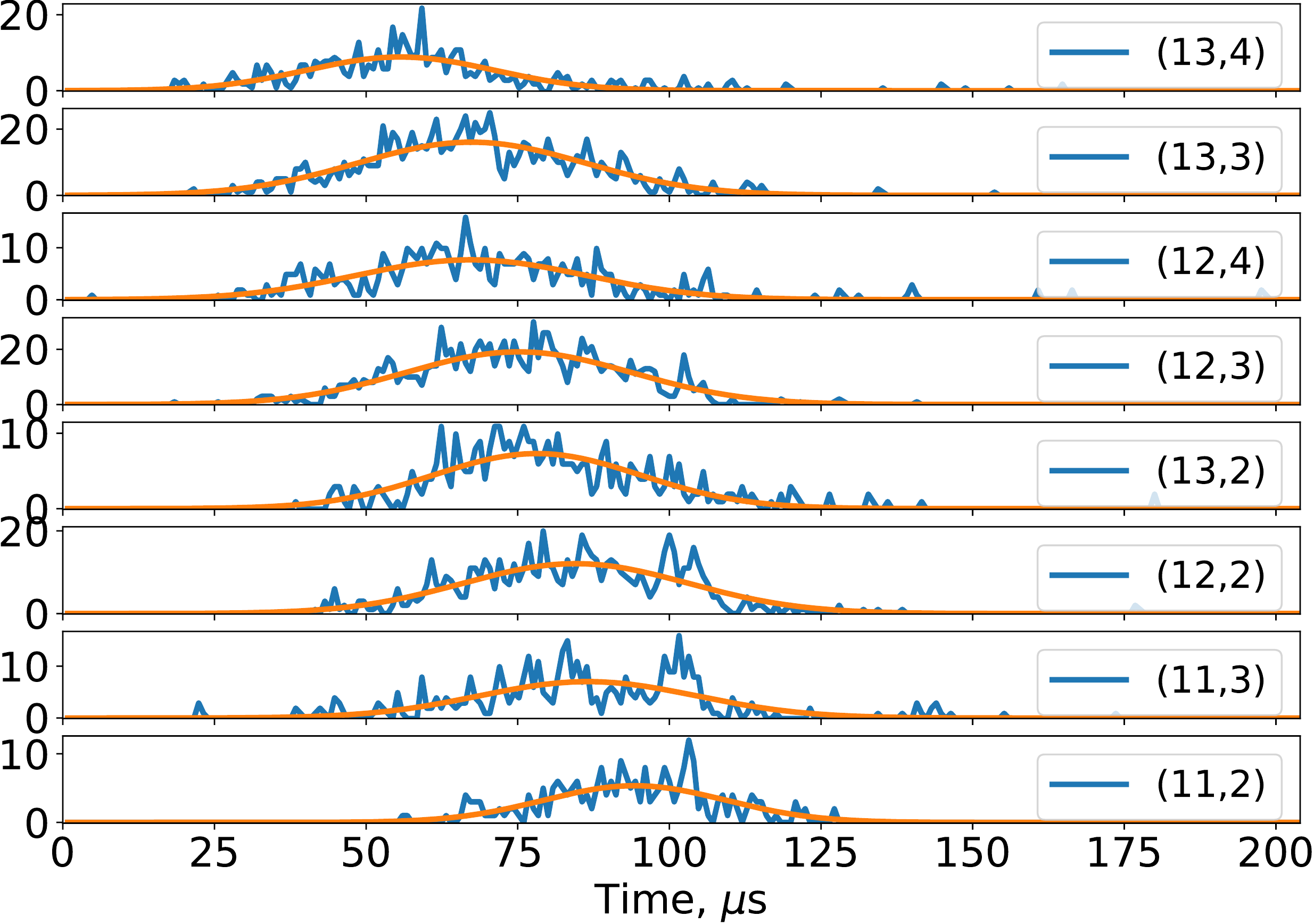}
	\fig{.56}{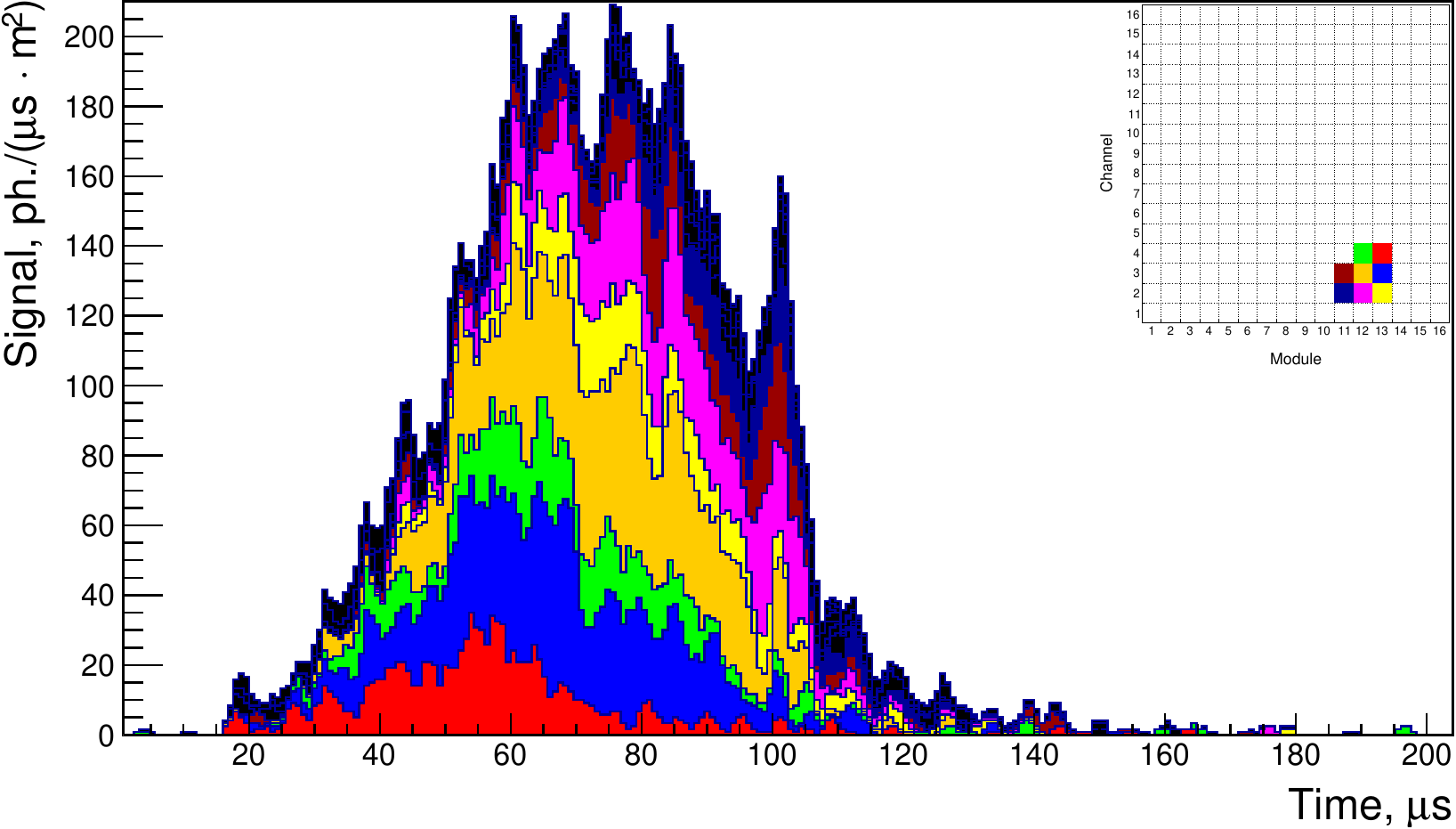}

	\caption{An expected TUS detector response to an EAS from a 1~ZeV proton
	arriving at the zenith angle $\theta\approx50^\circ$:
	ADC codes of the hit channels (left), the light curve and the focal
	surface (right).
	See the text for other details.
	}

	\label{fig:simulated}
\end{figure}

The left panel of Fig.~\ref{fig:simulated} shows waveforms of eight hit
channels that represent the detector response to an EAS from a primary
proton with the energy of $1~\text{ZeV}=10^{21}$~eV and the zenith angle
of $\approx50^\circ$.  For simplicity, we use an equal sensitivity of
all channels corresponding to the overall optical efficiency~0.14 and
the PMT gain~$10^6$. Zero background illumination is assumed.  An albedo
of the ground surface was set equal to~2\% since this is
the value of albedo of grass in the UV range of
320--390~nm~\cite{albedo}, and the TUS161003 event was registered above
a similar surface.  One can see a sequential movement of the peak of the
digital signal from one channel to another.\footnote{Channels are
indicated in the format (module number, channel number).} The right
panel of the figure represents the characteristic features of the light
curve calculated in accordance with Eq.~(\ref{eq:Sign}). One can see its
increase and subsequent abrupt decline due to the shower front hitting
the ground and a weak Cherenkov peak near the time stamp 100~$\mu$s.
Here,  the signals are filtered with the moving average
calculated over three time steps to minimize statistical fluctuations.
In the upper-right part of the panel, a channel map is shown, on which
the hit channels are highlighted with the same colours as in the
histogram.

\begin{figure}[ht!]
	\centerline{%
		\fig{.33}{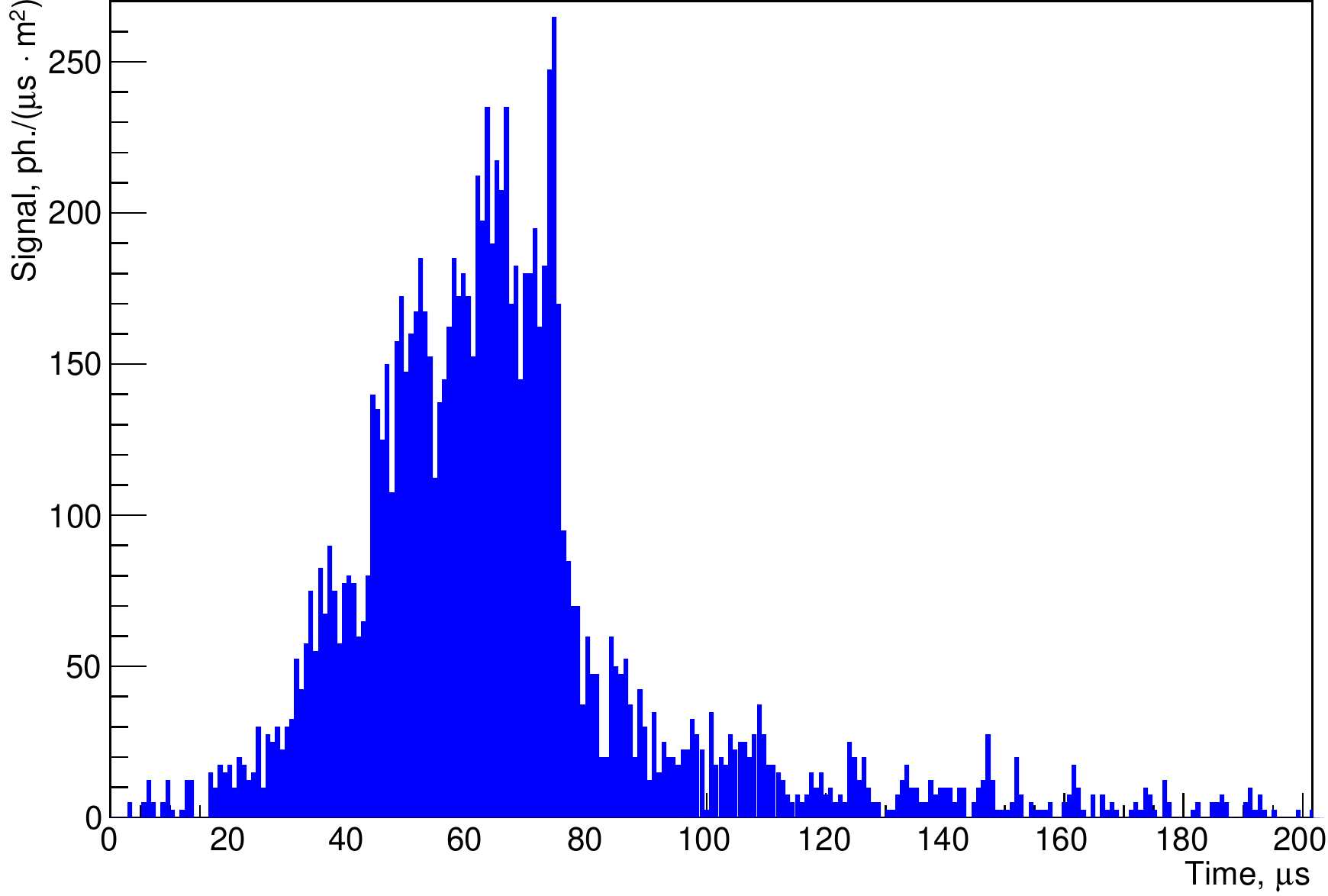}
		\fig{.33}{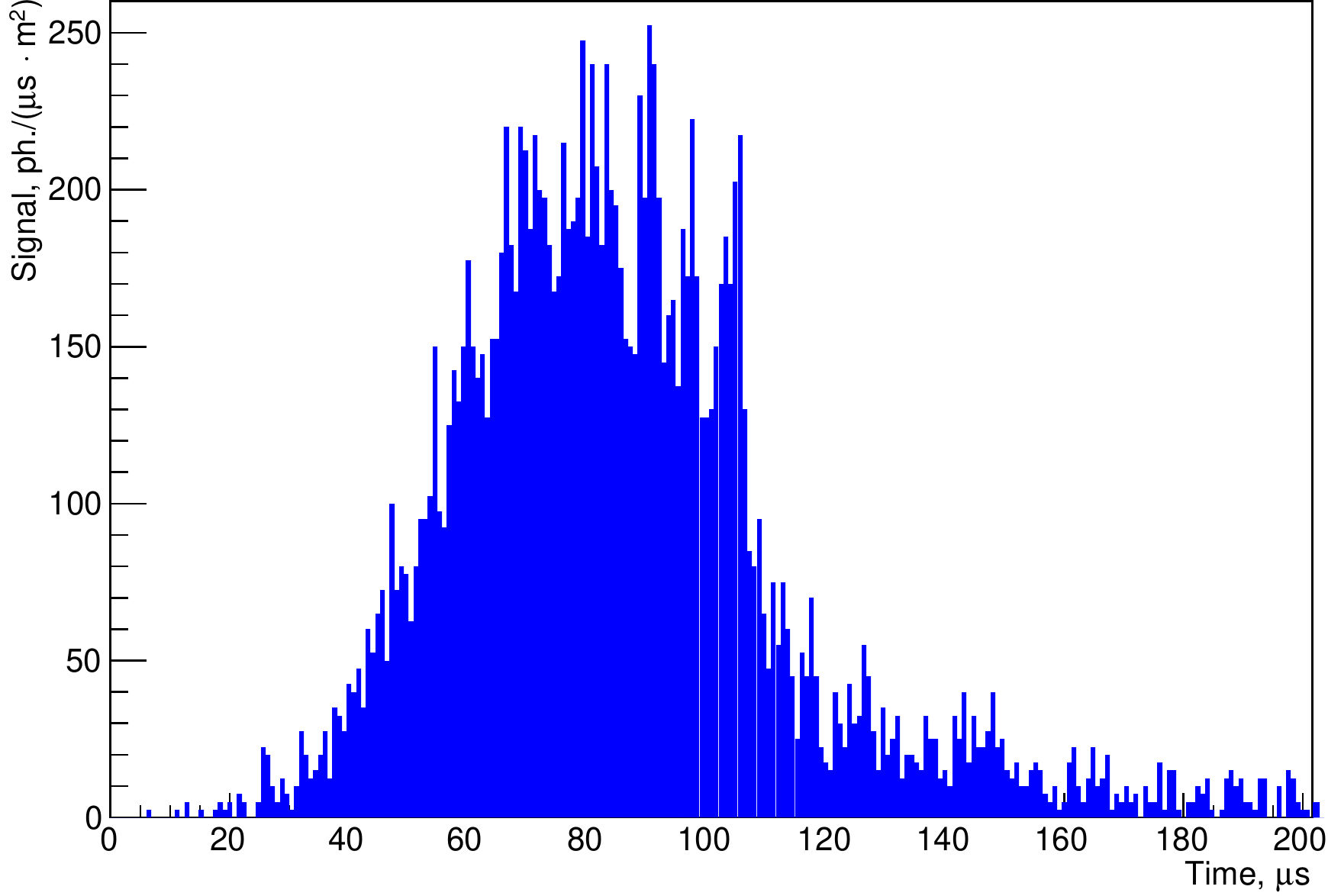}
		\fig{.33}{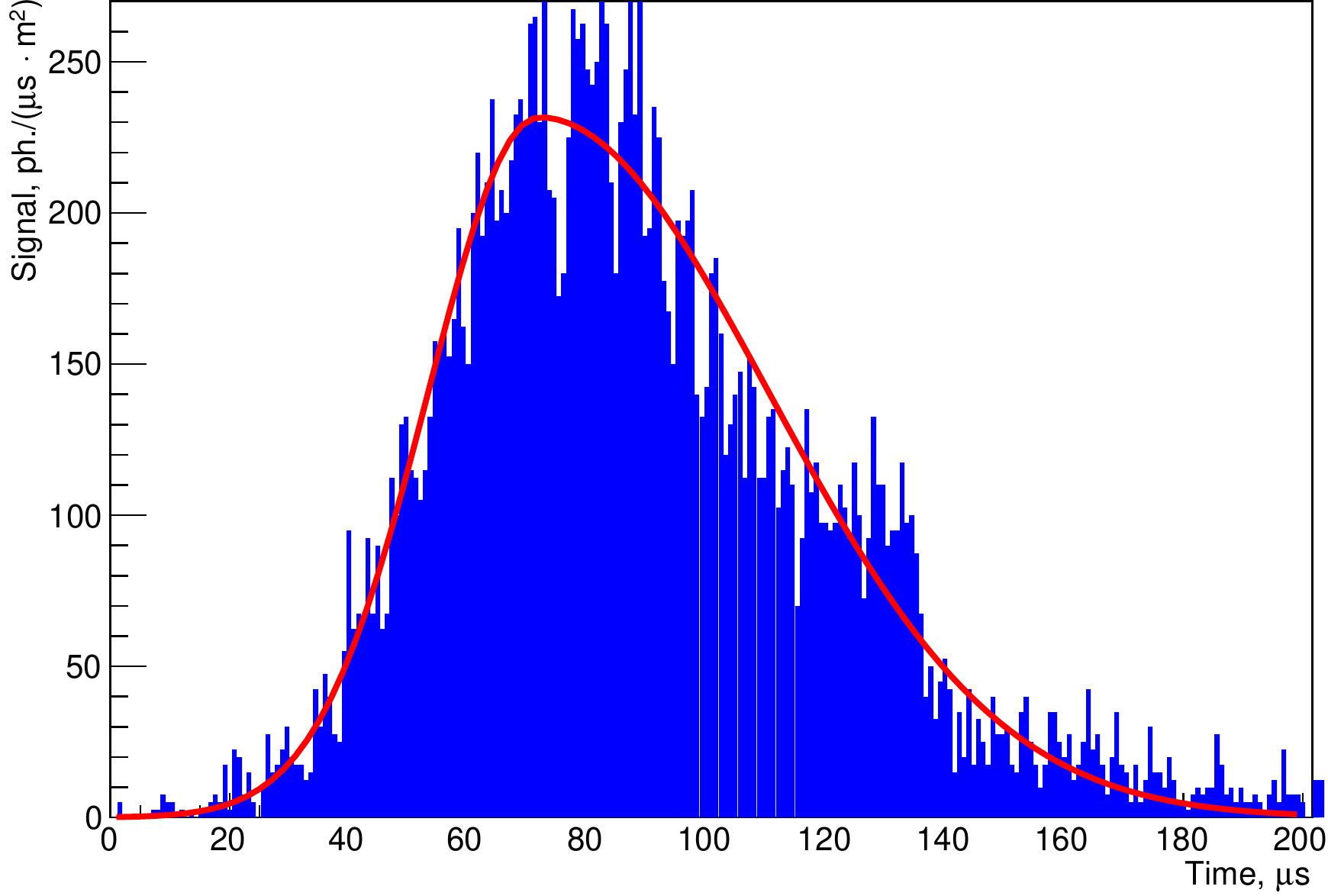}
		}

	\caption{Expected light curves of typical EASs originating from a
	1~ZeV proton arriving at the zenith angles $\theta=30^\circ, 45^\circ,
	60^\circ$ (from left to right) as seen by TUS, according to
	simulations with ESAF.  The red curve in the right panel shows a fit
	with an asymmetric Gaussian.  See the text for other details.
	}
	\label{fig:3simulated}
\end{figure}

Figure~\ref{fig:3simulated} presents examples of the expected response
of TUS to EASs generated by a primary proton with the energy of 1~ZeV
arriving at three different zenith angles $\theta=30^\circ, 45^\circ,
60^\circ$.  Simulations were performed with the same assumptions on the
sensitivity of channels, albedo and the background illumination as
above.  Arrival directions of EASs were chosen so that the signal was
located in approximately the same part of the focal surface as the
TUS161003 event. Five orders of scattering of light were taken into
account since an impact of higher orders was found to be negligible.


One can see typical asymmetric forms of the light curves depending on
the zenith angle of the primary proton, with the mean FDHM growing from
approximately 35~$\mu$s for $\theta=30^\circ$ to 70~$\mu$s for
$\theta=60^\circ$.  A sharp break of the signal at small zenith angles
is caused by the EAS front hitting the ground.  A pronounced peak from
the Cherenkov radiation is seen clearly for $\theta=30^\circ$
at $\sim80~\mu$s time.
The Cherenkov peak becomes less pronounced as the zenith angle grows,
and is rarely observed in simulations for $\theta\gtrsim45^\circ$ and
the surface albedo of~2\%.  The heavy ``tail'' of the signal, which can
be seen even for EASs with small zenith angles, is due to multiple
scattering of light in the atmosphere~\cite{Berat2010}. It becomes
particularly significant at large zenith angles.

\section{Phenomenology of the TUS161003 event}

The event was registered on 3rd October 2016 at 05:48:59~UTC, 00:48:59
local time.  The centre of the FOV of TUS at the moment of registration
was located at $44.08^\circ$N, $92.71^\circ$W above Minnesota, USA, in
approximately 100~km south-east from Minneapolis. The orbit height was
about 481~km above the sea level.

\begin{figure}[!ht]
	\centering
	\fig{.6}{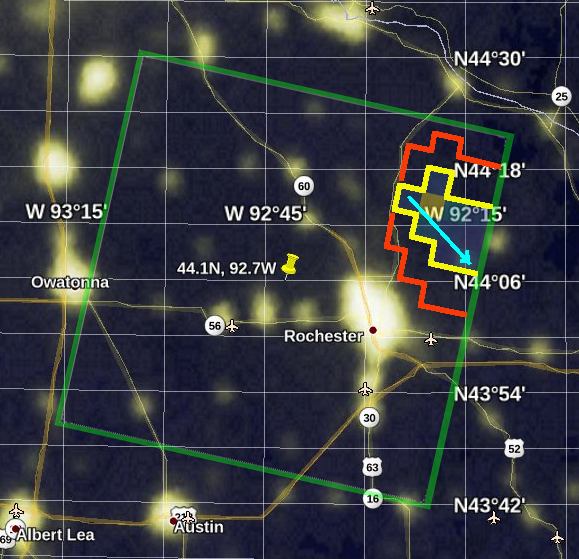}
	\caption{%
	Field of view of TUS (indicated by green lines)
	for the TUS161003 event on a
	Google Earth\protect\footnotemark{} map with the NASA data on the night
	Earth. Boundaries of the spot made by the hit channels is shown in yellow.
	The uncertainty of the position of these boundaries due to the discreteness
	of time stamps ($\pm1$~s) is shown in red. The cyan arrow indicates the
	direction of the apparent movement of the signal.
   }
	\label{fig:maps}
\end{figure}

Location of the FOV of TUS on the ground at the moment of registration
of the event is shown in Fig.~\ref{fig:maps}.
The direction of motion of the satellite was from NNE to SSW, parallel
to the left and right boundaries of the FOV shown in green.
The location is a part of the Richard J.~Dorer Memorial Hardwood State
Forest.  There are no big cities, airports or other obvious potential
sources of the signal below the hit channels.
The Red Wing Regional Airport ($44^\circ35'22''$N, $092^\circ29'06''$W)
and Rochester, the third-largest city of Minnesota, are located beyond
the ``error box'' of the location of the hit pixels.\footnote{%
The uncertainty in the position of the FOV of the detector is due
to the $\pm1$~s accuracy of the time stamps of the triggered events,
which results in approximately $\pm12$~km along the trajectory.
The coordinates of the satellite were known with a high accuracy
thus we take them as precise and neglect an uncertainty of the position
in the direction perpendicular to the direction of motion.}
One can also notice the Mississippi river, which is a major commercial
waterway, to the North-East of the FOV
of the detector, with the shape of the river being roughly parallel to
the spot made by the hit channels.\footnote{We thank the anonymous
referee for noticing this.}
We have studied the TUS data for correlations between triggered events,
especially those with similar waveforms, with the Mississippi River
and other major commercial waterways but did not find any.
A visit of our colleague to the area of registration did not reveal
any obvious bright sources of strobe lights though we cannot exclude
there were some at the moment of the observation.

\footnotetext{\url{https://www.google.com/earth/}}

\subsection{Description of the weather situation}
\label{sec:weather}

An analysis of conditions of observation was performed to
exclude possible atmospheric sources that could imitate an EAS signal.
The analysis was conducted by means of the US National Weather Service
dataset\footnote{\url{https://www.wpc.ncep.noaa.gov/index.shtml\#page=ovw}}
with the main conclusion that the atmosphere was clear at the
time of observation of the event without any noticeable clouds except
some small low-altitude ones. Details are provided below in Appendix~\ref{app:weather}..

Thunderstorm activity was studied in the region using the Vaisala Global
Lightning Dataset GLD360, a ground-based lightning location network with
a relatively high detection efficiency~\cite{Said2010,Vaisala2016}.  No
lightning strikes were registered within 930~km and during $\pm10$~s
from the event. This witnesses in favour of a non-thunderstorm origin of
the signal.

\subsection{Hit channels and their signals}

\begin{figure}[!ht]
	\centerline{%
		\fig{.5}{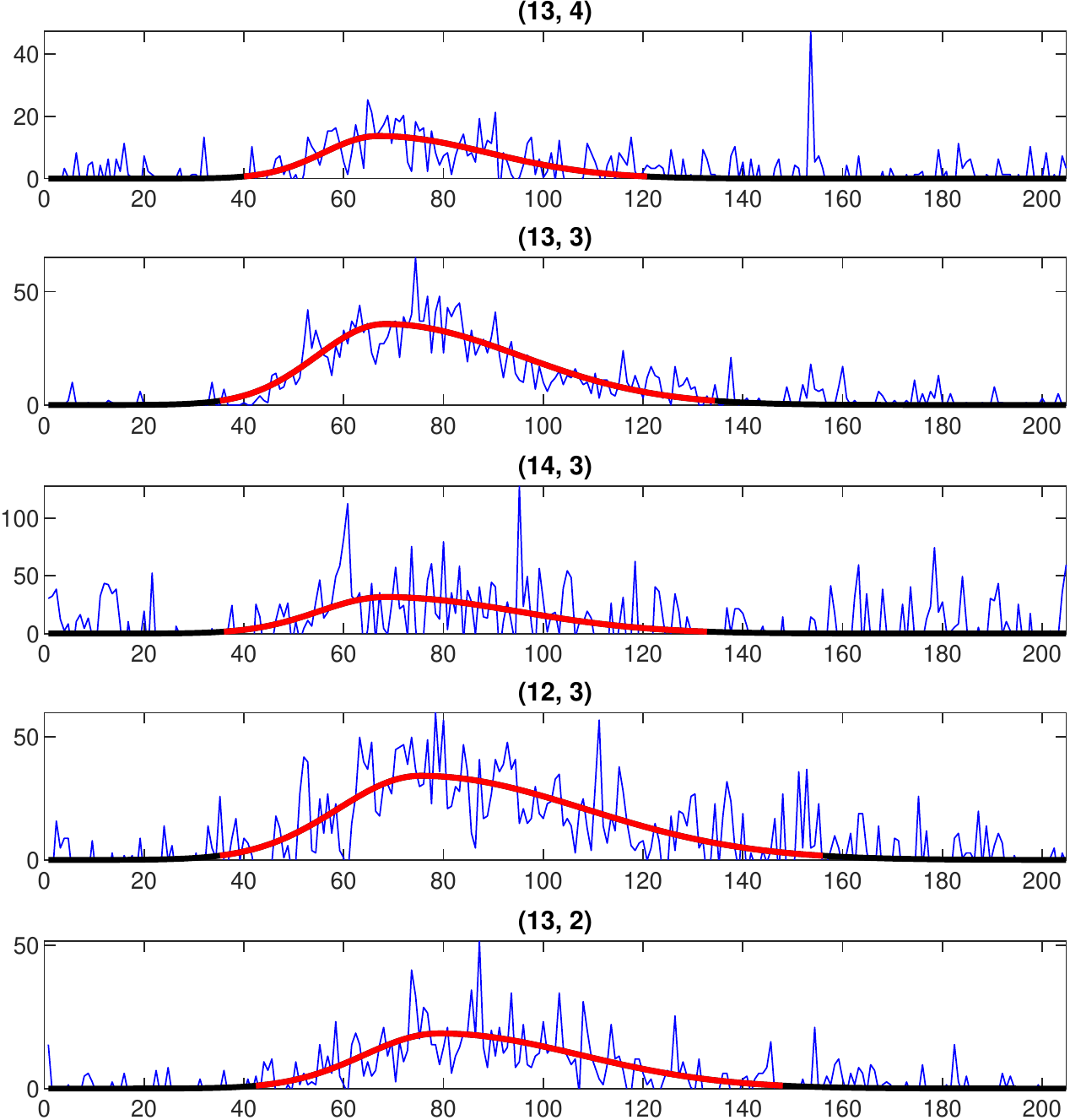}
		\fig{.5}{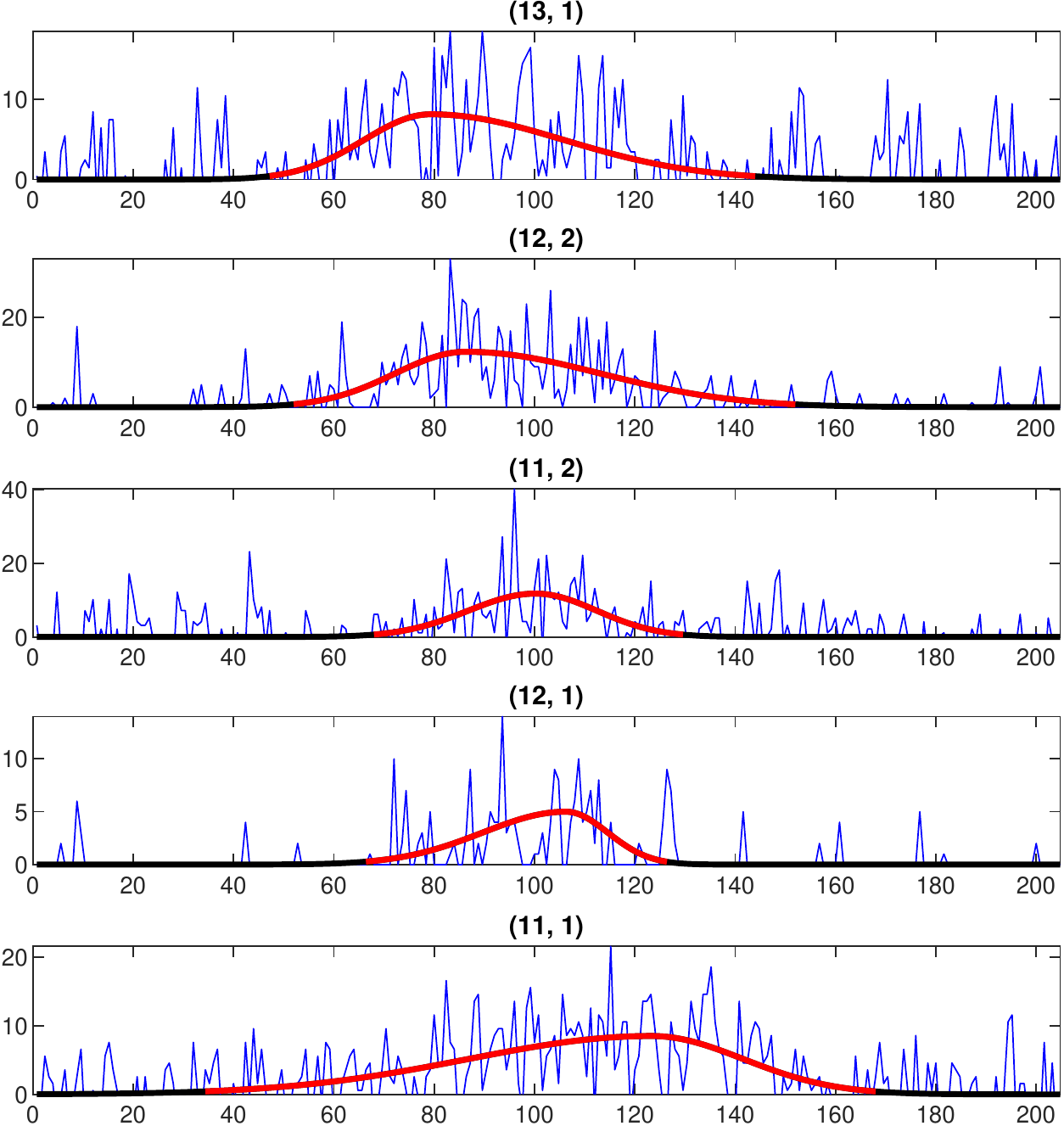}
   }
	\caption{Signals in the hit channels of the TUS161003 event
	with approximations.
	The average level of background illumination is extracted
	from the signals for clarity.
	Numbers in parenthesis above the panels indicate
	the module and channel numbers respectively.
	The $x$-axes represent time in~$\mu$s.
	}

	\label{fig:signals}
\end{figure}

An algorithm similar to the one developed for searching for EAS
candidates~\cite{2017arXiv170605369B} was employed for selecting hit
channels in the TUS161003 event.  A total of 10 hit channels were
selected. The respective waveforms are shown in Fig.~\ref{fig:signals}.
The signals are approximated by an asymmetric Gaussian function, i.e., a
smooth sewing of two Gaussian functions at the point of maximum.

The signals demonstrate a spatio-temporal dynamics similar to what is
expected from an EAS, see the left panel of Fig.~\ref{fig:simulated}.
The hit channels are grouped in an oblong spot, the shape of which might
be a ``convolution'' of two factors, namely, the asymmetric PSF of the
mirror and a linear track.  Times of the maximum of the signal (a peak)
in each channel have some shift from one channel to another.  It is
natural to interpret this as a movement of the signal in the FOV of the
detector.

Numerical characteristics of the signals in all hit channels are
presented in Table~\ref{tab:HitInfo}.  Shown are the time of the
peak~$t_\text{p}$ from the beginning of the record, the full duration at
half maximum (FDHM) and the amplitude~$I_\text{p}$ of the signal.  The
median FDHM equals approximately 48~$\mu$s, the time interval from the
first peak to the last one $\Delta t_\text{p}\approx 56~\mu$s.

\begin{table}[!ht]
	\caption{%
		Parameters of the signal in the hit channels: location of the
		pixels in the focal surface (module, channel numbers),
		position~$t_\text{p}$ of the peak from the beginning of the
		record,	the FDHM of the signal, and the peak
		amplitude~$I_\text{p}$ expressed in ph.~$\mu$s$^{-1}$~m$^{-2}$.
	}
	\begin{center}
	\begin{tabular}{|c|c|c|c|}
		\hline
		(md,ch) &	$t_\text{p},~\mu$s & FDHM,~$\mu$s & $I_\text{p}$ \\
		\hline
      (13,4)&	66.9&	38.9&   27.9\\
		(13,3)&	68.2&	48.0&   81.2\\
		(14,3)&	68.2&	47.0&	12.1\\
		(12,3)&	75.3&	58.3&	30.5\\
		(13,2)&	79.1&	51.2&	34.9\\
		(13,1)&	79.3&	46.9&	9.0\\
		(12,2)&	86.3&	48.3&	20.9\\
		(11,2)&100.2&	30.2&	11.8\\
		(12,1)&104.8&	40.9&	7.6\\
		(11,1)&123.3&	64.5&	10.6\\
    	\hline
	\end{tabular}
	\end{center}
	\label{tab:HitInfo}
 \end{table}

It is important to note that the rest of the channels demonstrated
a noise-like behaviour of the signal typical for a low background
illumination, similar to the behaviour of the signals shown in Fig.~\ref{fig:signals}
during the first $\sim40~\mu$s of the record.
The trigger rate was at the average level.
The signal could not be caused by a direct cosmic ray hitting the camera
because such events used to produce drastically different waveforms.
Such signals used to manifest themselves in a signal jumping in most cases to
the highest possible ADC counts during one time step (i.e., in less than
a microsecond) and then having exponential tails. The hit channels in
such events occupied adjacent pixels forming a line in the focal surface,
which made us call them ``track-like.''
Detailed discussions of this type of events with numerous examples can
be found in~\cite{Klimov2017,tus-jcap-2017,Zotov:uhecr2016,Zotov2018}.

\subsection{Light curve analysis}

\begin{figure}[!t]
   \centerline{\fig{.85}{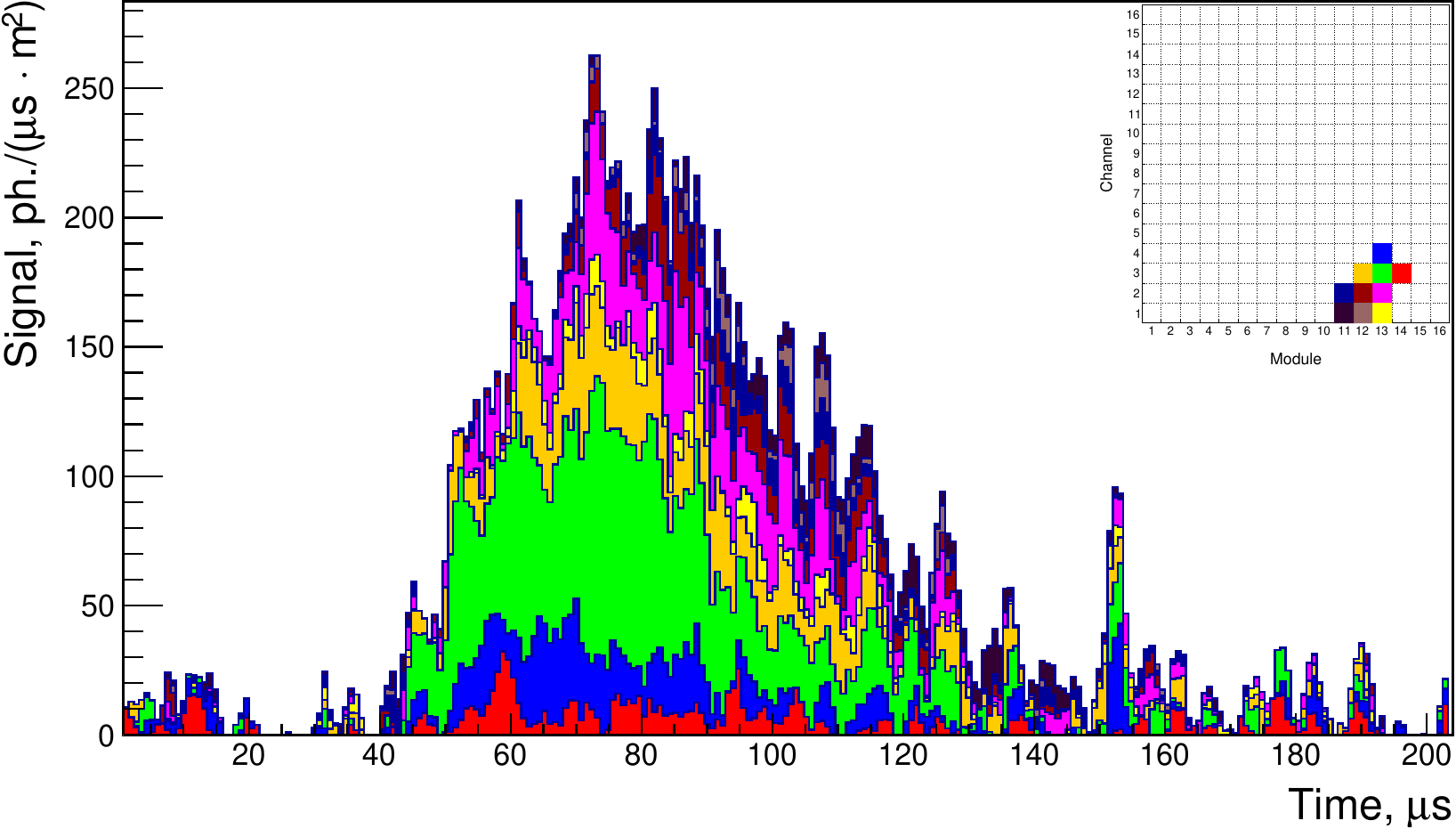}}
   \caption{The light curve of the TUS161003 event
   as the signal of the ten hit channels stacked together.
   The insert shows positions of the hit pixels in the focal
   surface.
   }
   \label{fig:event}
\end{figure}

Waveforms of the ten hit channels shown in Fig.~\ref{fig:signals} were
used to reconstruct the light curve of the event. The resulting light
curve and the hit channel map are shown in Fig.~\ref{fig:event}.  The
signal is smoothed with the moving average calculated for three time
steps to minimize statistical fluctuations.
 
The light curve is clearly asymmetric, which is most likely due to the
delayed signal as a result of multiple scattering of UV photons in the
atmosphere.  The following set of
four parameters was obtained by fitting the light curve with an
asymmetric Gaussian function with the coefficient of asymmetry defined
as a ratio of the attenuation time to the rise time:

\begin{itemize}\label{BiGauss}
	\item the amplitude of the signal at its maximum $I_\text{m} = 214$
	ph.~$\mu$s$^{-1}$~m$^{-2}$;
	\item the position of its maximum $t_\text{m} = 73.0$~$\mu$s;
	\item the full duration at half maximum FDHM $= 57.7$~$\mu$s;
	\item the coefficient of asymmetry $\alpha=2.36$.
\end{itemize}


To compare, the asymmetric Gaussian approximation of the light curve
obtained for a simulated EAS with the zenith angle $\theta=60^\circ$
shown in the right panel of Fig.~\ref{fig:3simulated} has the
coefficient of asymmetry $\alpha\approx2.1$. 

\section{Reconstruction of the event parameters}

Here we present results of a reconstruction of parameters of the
TUS161003 event assuming it was an air shower generated by an UHECR.
The kinematic and amplitude characteristics of the signal in the hit
channels will be used to estimate the arrival direction of a primary
particle.  The amplitude of the light curve will be employed to estimate
its energy.

\subsection{Arrival direction}\label{sec:Direst}

To reconstruct the arrival direction of a relativistic particle, one
needs to find parameters of its track in the focal plane (FP),
along which the image centre (a ``point'') moves at a constant velocity:
\begin{equation}
	x(t)=x_0+u_x(t-t_0), \quad y(t)=y_0+u_y(t-t_0),
	\label{eq:xy}
\end{equation}
where~$x(t)$ and~$y(t)$ are Cartesian coordinates of the image centre in
the FP at time~$t$, $(x_0,y_0)$ is a point on the track corresponding to
$t=t_0$, and $u_x, u_y$ are projections of the point velocity on
the axes of the local Cartesian coordinate system.%
\footnote{The $X$-axis coincides with direction of motion of the
detector, the $Y$-axis points along the modules, the $Z$-axis points to
the nadir, and the origin corresponds to the centre of the FP.}
However, when using approximation~(\ref{eq:xy}), one should keep in mind
that an image of a point source hardly ever occupies a single channel in
the FP because of the PSF, and the track is not exactly linear because
of the discrete structure of the focal surface.

A reconstruction of~$u_x, u_y$ allows one to calculate the arrival
direction of a light source, its azimuth and zenith angles:
\begin{equation}
	\phi=\arctan(u_y/u_x),\quad
	\theta \approx 2\arctan(\omega R/c) + \gamma\cos\Delta\phi,
	\label{eq:thetaphi}
\end{equation}
where $R$~is the distance between the detector and the source of light,
which can be assumed constant, $c$~is the speed of light,
$\omega=\sqrt{u_x^2+u_y^2}/f$ is the angular velocity of the signal in
the FP and~$f$ is the focal distance of the mirror, $\gamma$ is the
angle between the optical system axis and the direction to the source,
$\Delta\phi$ is the angle between ground projections of the line of sight
and the track velocity direction. 

The first term in the expression for $\theta$ in Eqs.~(\ref{eq:thetaphi})
corresponds to the centre of the FOV. For an off-axis event with an
angle~$\gamma$, one should take into account the correction to the
zenith angle, which means that the instantaneous velocity of the point
varies slightly along the track (see the second term).

An heuristic method of estimating parameters of a track was suggested
in~\cite{TkachevICRC2017}. The method is based on minimizing weighted
least squares for Eqs.~(\ref{eq:xy}), independently for~$x$ and~$y$.
Weights are chosen equal to ADC codes in hit channels.
An analysis that followed revealed a drawback of the method:
it happened to be sensitive to the choice of hit channels in case some
of them have a low signal. The method was modified to diminish the
effect. In the improved algorithm, which we called the Linear Track
Algorithm (LTA), one minimizes the sum
\begin{equation}
	\sum_{i} \sum_{k=k_1(i)}^{k_2(i)}W_i(t_k)\cdot(x_0+t_k u_x-X_i)^2
	\label{eq:LTA}
\end{equation}
over parameters $x_0$ and $u_x$, and a similar sum for~$y_0$ and~$u_y$.
Here, the index~$i$ runs over all hit channels in the event, $W_i(t_k)$
is a weight that correlates with the signal value at time~$t_k$,
and~$X_i$ and~$Y_i$ are coordinates of the centre of the $i$th pixel.

One of the key modifications aimed at reducing the influence of
statistical fluctuations of the ADC codes on the results of the
LTA-reconstruction is a search for moments~$t_k$ when the $i$th channel
is ``active.'' To solve the problem, the signal in each hit channel with
the base level subtracted was fitted with a Gaussian.  Then an
``activity window'' of the $i$th channel was defined as an interval
$[k_1(i),k_2(i)]$ such that the fit values exceed some threshold
value~$q_\mathrm{a}$.  Weights $W_i(t_k)$ are proportional to the $n$-th
power of the fit value in the $i$th channel at time~$t_k$. Parameters
$q_\mathrm{a}$ and $n$ control the LTA operation.  It was shown in a
dedicated study that the choice of the threshold $q_\mathrm{a}$ equal
to~30\% relative to the amplitude value makes the reconstruction
procedure more robust, and exponent~$n\approx1$ gives the smallest bias
in the reconstruction.

For the mean distance $R=480$~km, the off-axis correction angles $\gamma
= 4.1^\circ$ and $\Delta\phi = 74.8^\circ$, the LTA-reconstruction gave
the following results for the TUS161003 event: $\phi = 49.7^\circ$,
$\theta = 43.7^\circ$.

\begin{figure}[!t]
	\centerline{\fig{.49}{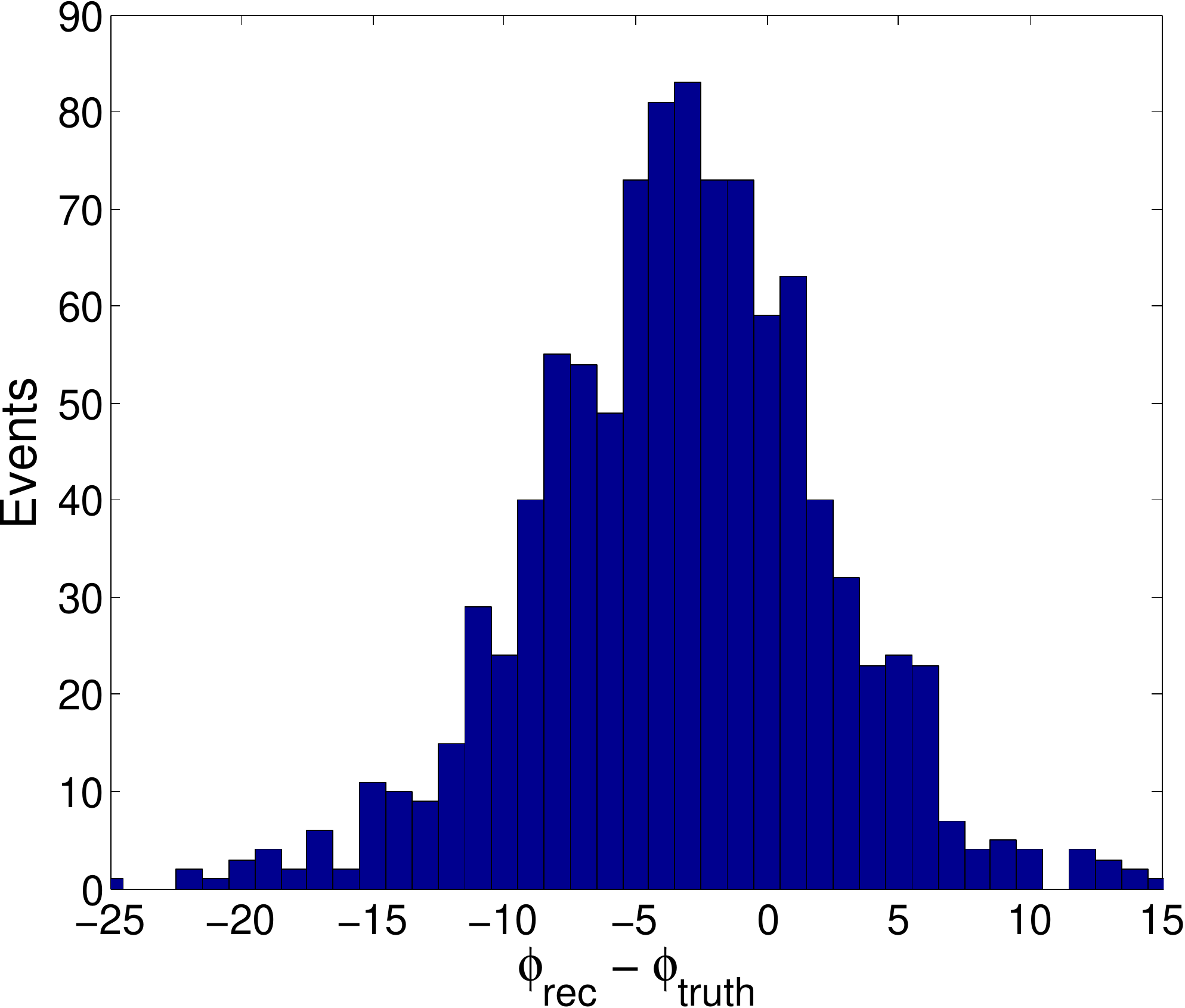}\fig{.48}{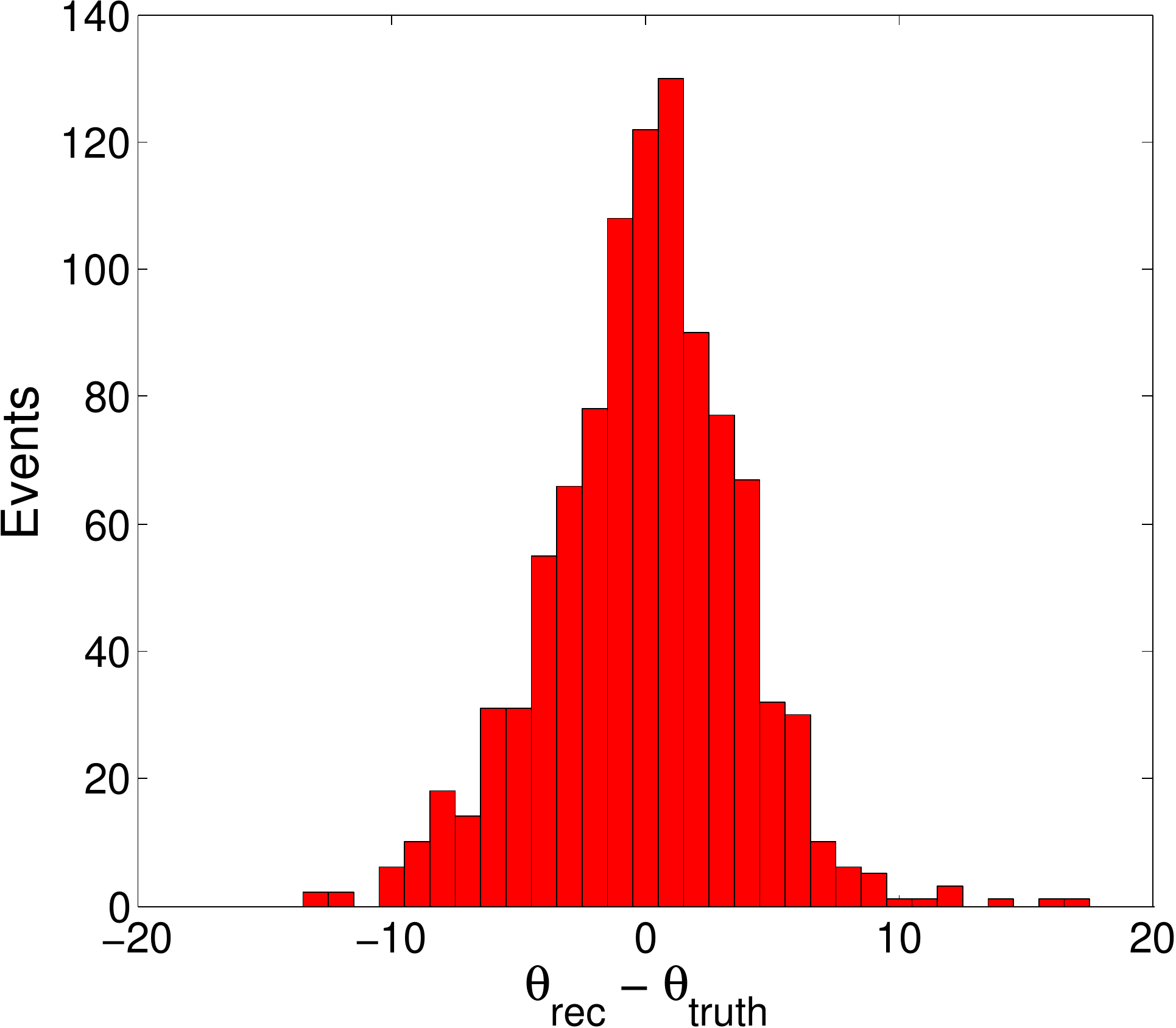}}

	\caption{%
		Distribution of errors of the reconstruction of
		the zenith and azimuth angles (the left and right panels
		respectively).
		}

	\label{fig:LTA-acc}
\end{figure}

Accuracy of the LTA method was studied with one thousand events
simulated in ESAF for $\phi\in[40^\circ, 60^\circ]$,
$\theta\in[35^\circ, 55^\circ]$.  A distribution of errors of the
reconstruction is presented in Fig.~\ref{fig:LTA-acc}.  One can see the
mean error of reconstructing the zenith angle is close to zero but the
reconstruction of the azimuth angle has a bias with the mean
error $\approx-3.5^\circ$.  A comparatively low accuracy of the
reconstruction is primarily due to the size of the pixels of the
photodetector and a relatively small number of the hit channels.  A work
to improve the accuracy of the reconstruction is currently in progress
and its results will be presented in a separate paper.

Assuming RMS-error estimated from simulated events as the uncertainty
of the reconstructed values, we obtain the following result for the
arrival direction of a hypothetical source of the TUS161003 event:
\[
\phi_\mathrm{rec} = \left(50\pm10\right)^\circ,\quad
\theta_\mathrm{rec} = \left(44\pm4\right)^\circ.
\]

\subsection{Energy estimation}
\label{sec:Eest}

It is well known that the number of photons in an EAS maximum is
proportional to the number of electrons and, thus, the energy of a
primary particle.  This allows one to estimate the energy~$E$ of a
cosmic ray that caused the EAS basing on the amplitude of the
light curve obtained after an absolute calibration of the detector.

Taking into account only direct fluorescent radiation, the amplitude of
the signal at the entrance pupil can be written as
\begin{equation}\label{eq:IvsE}
	I_\mathrm{fl} = 
	N_\mathrm{max}\frac{c\, \eta_\mathrm{atm}Y_\mathrm{fl}}
	{4\pi R^2} \frac{\cos\gamma}{1+\cos\theta},
\end{equation}
where $N_\mathrm{max}$ is the number of electrons (and positrons) in the
maximum, which is proportional to the energy of a primary particle
$N_\mathrm{max}\approx E/1.45$~GeV \cite{Berat2010}, $Y_\mathrm{fl}$ is
the fluorescence yield and $\eta_\mathrm{atm}$ is the atmosphere
transmittance.

The fluorescence yield $Y_\mathrm{fl}$ (the number of photons per
charged particle per meter) is approximately constant at altitudes from
5~km to 15~km ($Y_\mathrm{fl}\approx 5$~ph./m). The atmosphere
transmittance~$\eta_\mathrm{atm}$ depends on the altitude of the EAS
maximum. If we take into account the Rayleigh scattering only, the mean
transmittance in the wavelength 320--400~nm in the nadir direction
increases from~0.6 to~0.85 as the altitude changes from 5~km to 10~km
(and is greater than~0.9 above 14~km).

As a result, an UHECR with an energy of 1~ZeV and a zenith angle
$\theta_\mathrm{rec} = 44^\circ$ generates direct fluorescence
illumination of the order of 150--180~ph.~$\mu$s$^{-1}$~m$^{-2}$ for the
same altitude range.  However, as noted above, the signal recorded by
the TUS detector includes not only direct fluorescent radiation, but
also single- and multiple-scattered components of both fluorescent and
Cherenkov light.  Since the amplitude of the direct fluorescent
radiation is $\sim$70--80\% of the amplitude of the total signal at the
entrance pupil of the detector (see, e.g.,~\cite{Berat2010}), one
can expect the intensity at the maximum at the level of
200--240~ph.~$\mu$s$^{-1}$~m$^{-2}$, i.e., similar to that for the
TUS161003 event.

To make the energy estimation more accurate, we compared amplitudes of
light curves of the events simulated in ESAF with that of the TUS161003
event.  We simulated the detector response to EASs with zenith angles
$\theta=35^\circ\dots55^\circ$ and azimuth angles
$\phi=40^\circ\dots60^\circ$ located in the focal plane similar to the
TUS161003 event. Energies of primary particles (protons) varied in the
range from 200~EeV to 1000~EeV with a step of 200~EeV.
One thousand events were simulated for each energy bin.
As expected, the amplitude of the signal was found to be proportional to
the energy of a primary particle and reaches
212--230~ph.~$\mu$s$^{-1}$~m$^{-2}$ at 1~ZeV, see
Table~\ref{tab:EvsS}.

\begin{table}[!ht]
	\caption{
		Dependence of the light curve maximum on the primary
		energy of a proton arriving at zenith angles
		$\theta=35^\circ\dots55^\circ$.}

	\begin{center}
   	\begin{tabular}{|l|c|c|c|c|c|}
			\hline
			Energy, EeV & 200 & 400 & 600 & 800 & 1000 \\
			\hline
      	$\langle I_\mathrm{m}\rangle \pm\sigma_I$,~ph.~$\mu$s$^{-1}$~m$^{-2}$
				& $44\pm2$ & $88\pm3$ & $132\pm5$ & $176\pm7$ & $221\pm9$ \\
    		\hline
		\end{tabular}
	\end{center}
	\label{tab:EvsS}
\end{table}

One needs to take into account two factors when evaluating the accuracy
of energy estimations. These are (i)~fluctuations of amplitudes of
simulated events and (ii) an accuracy of estimating the amplitude of a
real light curve. The first factor is presented in Table~\ref{tab:EvsS}
as~$\sigma_I$ and leads to an error of $\sim5\%$ in the ZeV region.  The
second factor depends on the accuracy of the sensitivity estimation,
see Appendix~A below.  It is also influenced by a
possible loss of a part of the signal hidden in fluctuations of the
background illumination. This loss results in a small systematic
underestimation of the energy of a primary particle.

Thus, the lower estimate of the energy of the TUS161003 event in the
assumption of its UHECR origin can be presented as $E \gtrsim
10^{21}$~eV.

\section{Discussion}
\label{sec:discussion}

First of all, we conclude that an apparent relativistic movement of a
source of UV radiation was registered in the TUS161003 event. The
angular velocity of the image on the focal plane is $\omega=262$~rad/s,
which corresponds to the linear velocity around $0.43c$ of an object
moving at the distance of $\sim500$~km from the detector perpendicular
to the line of sight (an apparent speed). This justifies the application
of the kinematic LTA method for reconstructing the arrival direction of
the source as described in Section~\ref{sec:Direst}.  Results of this
reconstruction demonstrate that the direction of movement is toward the
Earth surface with a zenith angle near $44^\circ$.  The most natural
candidates for being a source of UV radiation spreading downward to the
Earth are an extensive air shower initiated by an UHECR particle or a
beam of light, e.g., a laser shot from an airplane or a satellite.


The next important parameter of the track is the maximum luminosity of
the light curve, which is used to estimate the energy of a primary
particle generating the air shower.  It was shown in
Section~\ref{sec:Eest} that the energy of the TUS161003 event the
assuming its UHECR origin is of the order of $\gtrsim1$~ZeV.  This leads
to a difficulty in interpreting the event as a ``traditional'' cosmic
ray due to a very steep CR spectrum beyond 50~EeV measured by the main
ground-based experiments---the Pierre Auger
Observatory~\cite{Auger-spectrum-2017} and the Telescope
Array~\cite{TA-spectrum-ICRC-2017}.

Taking into account a limited exposure of the TUS experiment and the
fact that the flux of ZeV CRs is expected to be from two to four orders
of magnitude lower than that around 100~EeV, a chance TUS registered
such an extremely energetic cosmic particle is very low (order of
$10^{-3}\text{--}10^{-5}$) and, therefore, has to be excluded in the
first instance.

Another important phenomenological feature of the event is that the
maximum of the emission takes place very high in the atmosphere
assuming its UHECR origin.  This
conclusion can be made if we compare the light curve of a simulated
1~ZeV air shower shown in Fig.~\ref{fig:simulated} with that of the
TUS161003 event shown in Fig.~\ref{fig:event}.  It is clearly seen that
the registered light curve does not have a steep break expected to take
place when the shower front reaches the ground but gradually decays, nor
does it demonstrate a Cherenkov peak following the maximum of the light
curve in around 20--30~$\mu$s.  The time interval between the maximum of
the light curve and the moment when the signal decays to the background
level is $\sim$60~$\mu$s, which geometrically corresponds to the
altitude of $\sim$7.5~km.  This allows us to estimate the slant depth of
the shower maximum as $\sim550$~g/cm$^2$.

A high altitude of the light curve maximum can be also obtained if we
consider a sharp peak near 150~$\mu$s as a Cherenkov one, see
Fig.~\ref{fig:event}. It has a delay from the light curve maximum
$\sim70~\mu$s, and this means an altitude $\sim$8.5~km,
which corresponds to the slant depth $\sim480$~g/cm$^2$
This number cannot be explained upon the assumption of an
ordinary EAS with an energy of 100--1000~EeV and allows one to rule out
the proton origin of the air shower since its maximum at energies around
1~ZeV should be much closer to the ground level. 

\begin{figure}[!t]

	\centerline{\fig{.5}{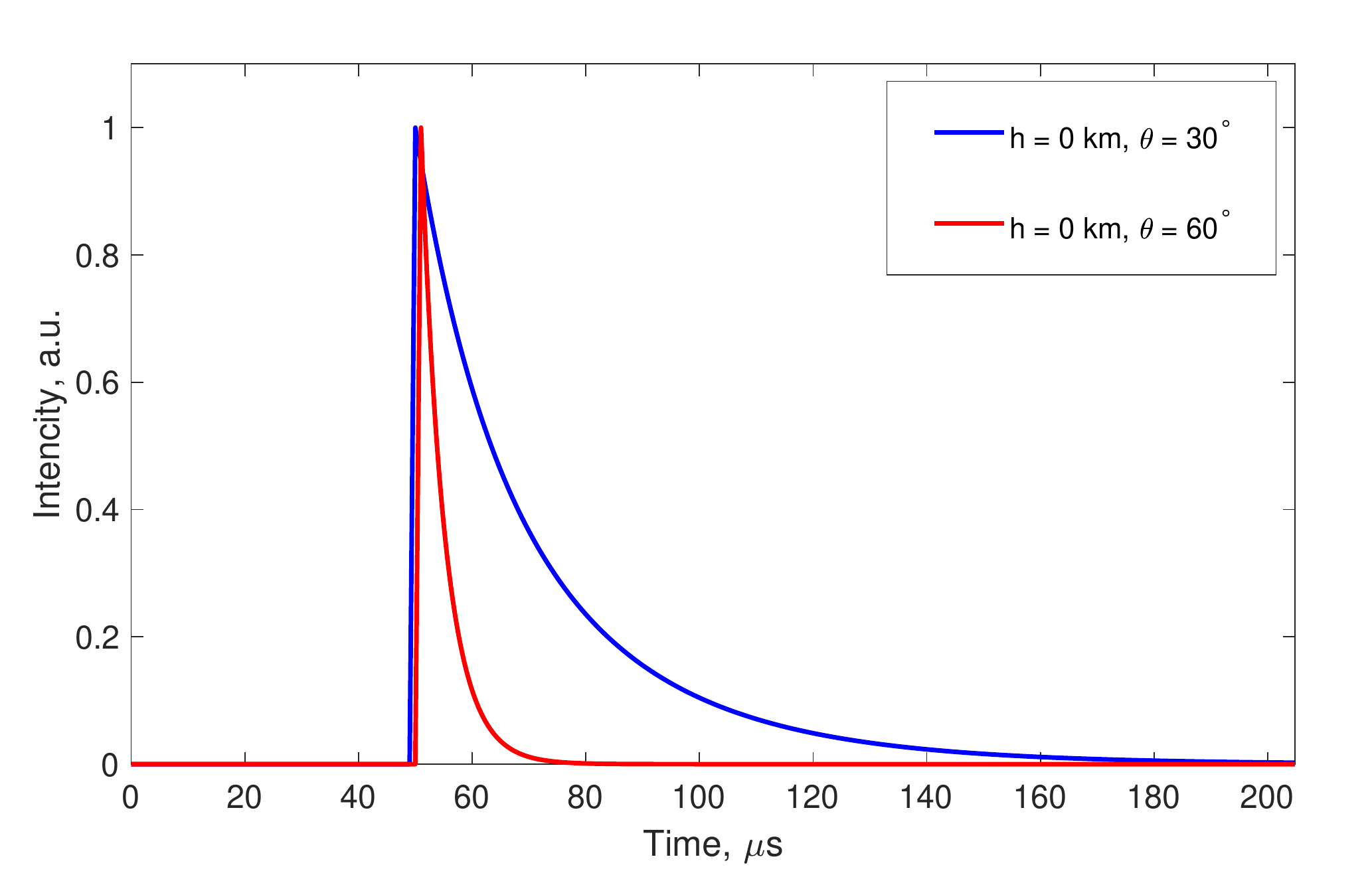} \fig{.5}{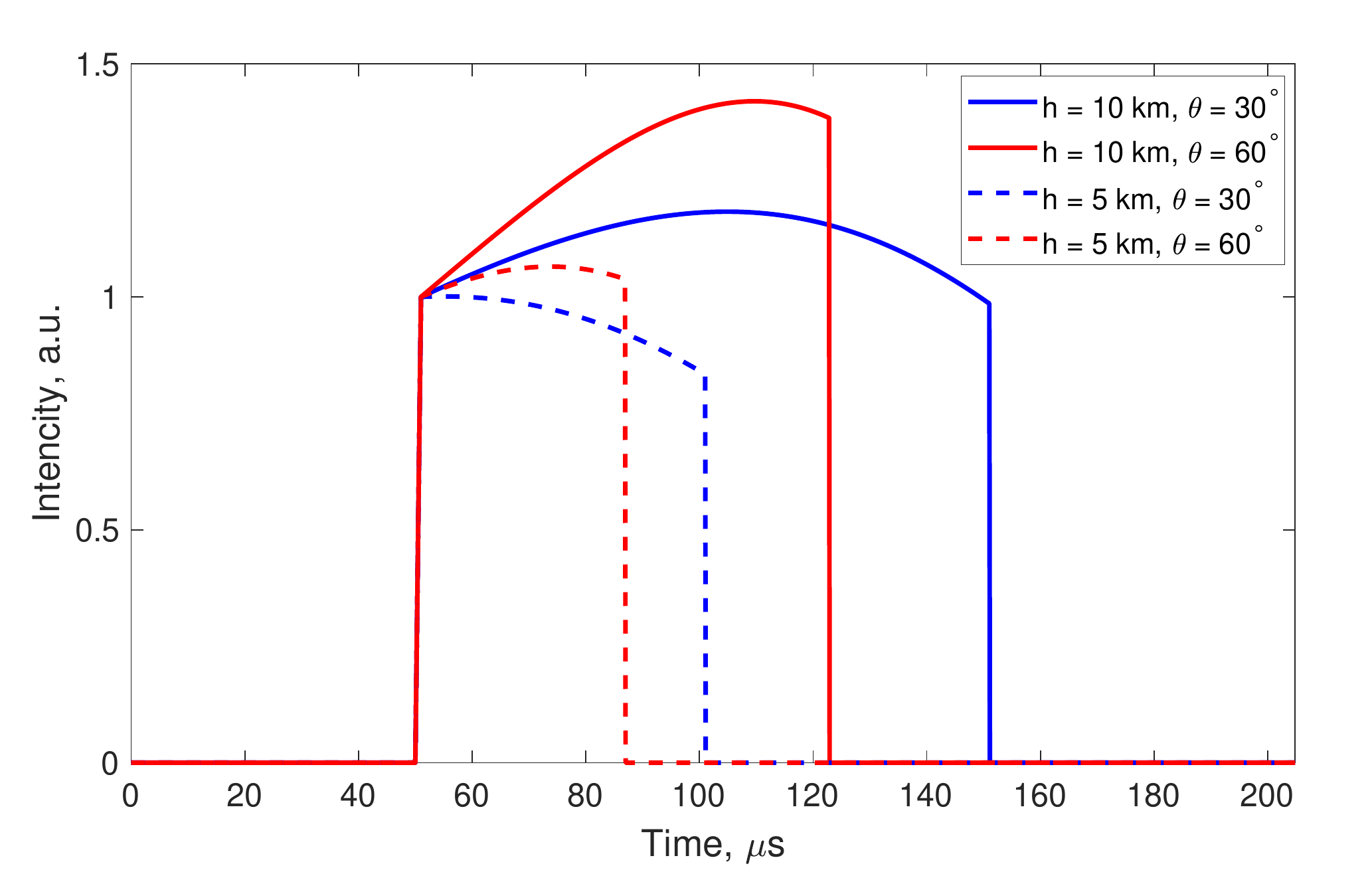}}

	\caption{Qualitative behaviour of light curves from scattering laser
	beams. Left panel: two upward going lasers with $\theta=30^\circ$
	(blue) and $60^\circ$ (red). Right panel: downward lasers (with the
	same zenith angle) starting from two different heights, $h=10$~km and
	5~km.  The amplitude of the light curves is shown in arbitrary units
	normalized to the value at the initial moment of time at 50~$\mu$s.}

	\label{fig:laser}
\end{figure}

The above energy considerations and the fact the TUS161003 event was
registered in a populated area, makes one
consider an anthropogenic origin of the event more closely.
One of the possibilities consistent
with the downward movement is a laser shooting from the height of a few
kilometers, for example, from an airplane.  Simple simulations show the
observed light curve cannot be explained by a laser lidar operating in a
pulse mode since the light curve of an upward-going laser beam should
have an exponential decay while the light curve of a downward-going laser
beam should exhibit a sharp cut-off as soon as the beam hits the ground, see
Fig.~\ref{fig:laser}.
However, multi-scattering can smoothen the cut-off resulting in a kind
of a tail of the light curve.
Anyway, the variety of devices
employing lasers is rich and one cannot exclude a signal similar
to the TUS161003 event can be produced by a combination of laser shots.

\begin{figure}[!ht]
    \centering
    \fig{.5}{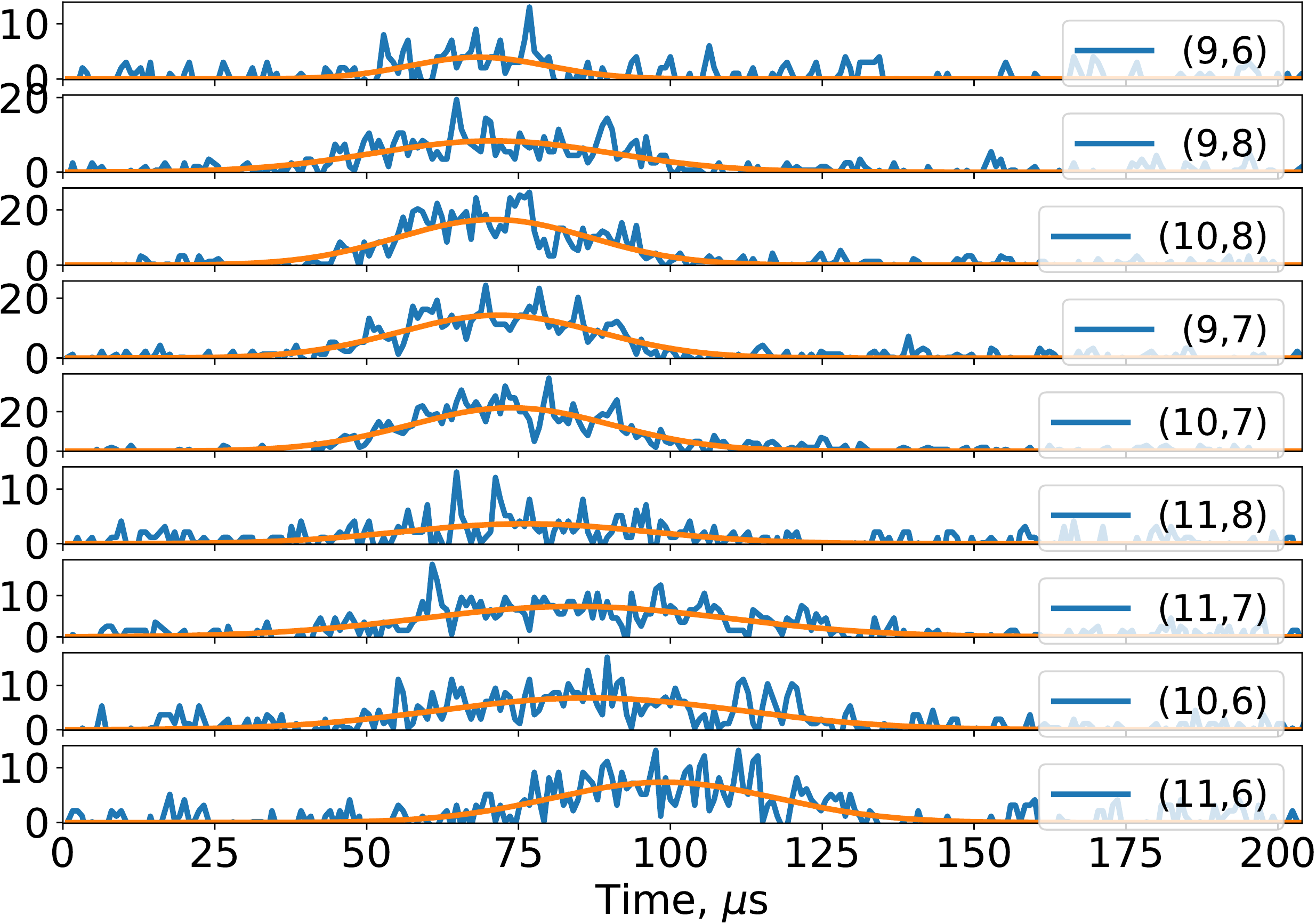}
    \caption{Waveforms of hit channels of a signal combined of
    two nearby on-ground flashes, see the text for details.}
    \label{fig:flashers}
\end{figure}

Another possibility is an ensemble of light impulses produced, e.g.,
by Xenon flashers. Simulations have revealed that a specially crafted
pair of short but bright light flashes with a duration of the order of
40--50~$\mu$s, located in 7--10~km from each other,
separated by $\sim30\text{--}40~\mu$s in time, and with the first flash
being 2--4 times brighter than the later one, can produce a signal with the
light curve and kinematics of the signal similar to the TUS161003 event.
Waveforms in hit channels of such an event are demonstrated
in Fig.~\ref{fig:flashers}.
A similarity with the signals shown in Figs.~\ref{fig:simulated}
and~\ref{fig:signals} is evident.
An application of the LTA algorithm resulted in a ``zenith angle''
of the order of 35$^\circ$ in this case.

An emission from a pair of such flashes seems to be the most
straightforward explanation of the TUS161003
signal taking into account the event was registered above a populated
area, providing Xenon or other flashers emit enough light in
the UV range.
Still there are some caveats with this.
First, these should be very short flashes bright in UV, with certain duration,
mutual delay in time and separation in distance.
Next, assuming these are random coincidences between flashes, it is
natural to expect that in a half of cases the power of the first flash
should be greater than the power of the second one, and the situation
must be opposite for the other half of the events.
As we have shown above, the signal in the TUS161003 event had
a steep front and a much longer tail, which implies the brighter flash
comes first. The other five events with similar waveforms and kinematics
of the signal, mentioned in the Introduction, demonstrated similar
asymmetric shapes of the signals and the light curves with a
steep front coming first, thus also implying a brighter flash at
the beginning of the event.
This makes a random coincidence of all these factors less plausible.

Other possible sources of air showers in the atmosphere are neutrinos,
energetic gamma-rays or heavier nuclei.
No astrophysical neutrinos or gamma-rays with energies above a few PeV
have been registered thus far but their existence is not ruled out by
modern theories.  It is important though that neutrino-induced
showers develop even deeper in the atmosphere due to the small
cross-section, and it is more likely to observe an upward-going EAS from
an Earth-skimming neutrino, which contradicts the reconstructed
direction of movement.  One can expect that a heavier primary particle
can solve this problem, but simulations show that the observed maximum
is much higher than can be expected even for an iron nuclei.

Relativistic dust grains suggest an interesting explanation of an
extreme-energy event developing high in the atmosphere. They were
considered long ago by Spitzer~\cite{Spitzer1949} and later by
Hayakawa~\cite{Hayakawa1972} as possible sources of UHECRs of the
highest energy.  According to simulations~\cite{PhysRevD.61.087302,
2001AIPC..566...57K}, an EAS initiated by a massive relativistic dust
grain develops in the atmosphere at slant depths
$\sim200$--400~g/cm$^2$, and this seemingly agrees with the
parameters of the TUS161003 event and can also explain why ground-based
fluorescent detectors, which observe the lower layer of the atmosphere,
do not register such events.
A more detailed analysis of this hypothesis and the probability of
registering dust grains by TUS will be addressed elsewhere.

\section{Conclusions}

The orbital telescope TUS on board the Lomonosov satellite is the first
instrument in space with the primary goal to explore the possibility of
registering UHECRs by their UV tracks in the atmosphere. The presented
analysis of the data obtained during the experiment allowed us to
interpret the TUS161003 event as a possible signal from an extensive air
shower. Assuming the cosmic ray origin of the cascade, our
reconstruction of the zenith angle gives
$\approx40^\circ\text{--}48^\circ$, which agrees with results of
simulations but the estimate of the signal amplitude indicates the
energy of a primary particle of the order of $10^{21}$~eV, which is
incompatible with the cosmic ray energy spectrum obtained with the
ground-based experiments.

Basing on the first estimation of the TUS exposure
$\sim1200\text{--}1400~\text{km}^2~\text{sr~yr}$ and the pre-flight
estimations of its sensitivity to UHECRs, it would be plausible to
expect measuring one event in the energy range of $10^{20}$~eV.  Due to
a problem which occurred after the launch, it is not possible to assume
the nominal sensitivity, and a new estimation of the TUS sensitivity was
obtained, see Appendix~\ref{Calibr}.  The energy evaluation for the
TUS161003 event based on these new estimations provides a value of one
order of magnitude higher energy than could be expected.  This can be
interpreted in several ways, among them the event does not have an UHECR
origin, see some hypotheses discussed in Section~\ref{sec:discussion}.

In conclusion, despite the fact that the origin of this event can not be
unambiguously determined at present, this detection proves the
capability of TUS to detect and trigger on light signals with an apparent
motion and the light curve similar to what is expected from EASs.
The most plausible interpretation of the event impies its
anthropogenic nature but we cannot totally rule out all possible astrophysical
sources including relativistic dust grains or even more exotic objects.
A further analysis of the
TUS161003 event and some other events registered during the experiment
is aimed to resolve the puzzle.  In the meanwhile, the Mini-EUSO
detector~\cite{mini-Capel-2018} with its larger FOV, even though with
less sensitivity than TUS, could search for the similar type of events
to help understanding their origin.

We believe this measurement is important for the future orbital
missions aimed for registering UHECRs from space. Being a pathfinder
with a relatively low sensitivity, narrow field of view and only the
5-km spatial resolution, TUS has proved the possibility of observing
EAS-like events from space and highlighted necessary improvements for
the next-generation missions like KLYPVE-EUSO
(K-EUSO)~\cite{K-EUSO2017} and POEMMA~\cite{POEMMA}.

\acknowledgments

We highly appreciate the invaluable help of Francesco~Fenu
with the implementation of TUS for ESAF and other issues
related to the analysis of the event.
We thank Gabriel~Chiritoi and Anthony~Salsi for their
contribution in the study at its early stage and
Roberto~Cremonini for the help with the cloud coverage analysis.
We acknowledge multiple insightful comments by the anonymous referee.
We thank Vaisala Inc.\ company for providing the relevant data on
lightning strikes.
The data used in Section~\ref{sec:weather} were retrieved from
the official NOAA National Weather Service
(\url{https://www.nco.ncep.noaa.gov/}). 
Satellite data were retrieved from the Satellite Data Services archive
(SSEC, \url{https://www.ssec.wisc.edu/}), and from the NASA LAADS DAC
archive (\url{https://ladsweb.modaps.eosdis.nasa.gov/}).
The work was done with
partial financial support from the State Space Corporation ROSCOSMOS,
M.V. Lomonosov Moscow State University through its “Prospects for
Development” program and the Russian Foundation for Basic Research grant
No.\ 16-29-13065 and 15-02-05498/17-a.  The Italian group acknowledges
financial contribution from the agreement ASI-INAF n.2017-14-H.O.

\appendix
\section{Photodetector calibration}
\label{Calibr}

The absolute calibration of PMTs is a crucial point for reconstructing
the number of in-falling photons and thus the energy and the arrival
direction of a primary particle of an observed EAS. The initial
calibration was performed before the launch of the Lomonosov satellite
at the wavelength of 375~nm~\cite{SSR}.  However, an emergency situation
occurred during the first orbits after the detector was switched on.
Namely, the system of an automatic PMT gain control did not work
properly on the day side of the orbits, so that PMTs used to operate
along the whole orbits at the highest voltage intended for nocturnal
segments only. The behaviour of the gain control system was fixed later
but a number of PMTs were damaged and gains of other PMTs changed.  As
a result, the absolute sensitivity of the channels turned out to be
known only with a big uncertainty.  A new method of the channel
calibration was developed to address the problem.  The method is a
combination of the PMT calibration based on an analysis of the
stationary background glow and an estimation of the relative sensitivity
of channels employing events registered in the slow (``Meteor'') mode.

\begin{figure}[!t]
	\centerline{\fig{0.48}{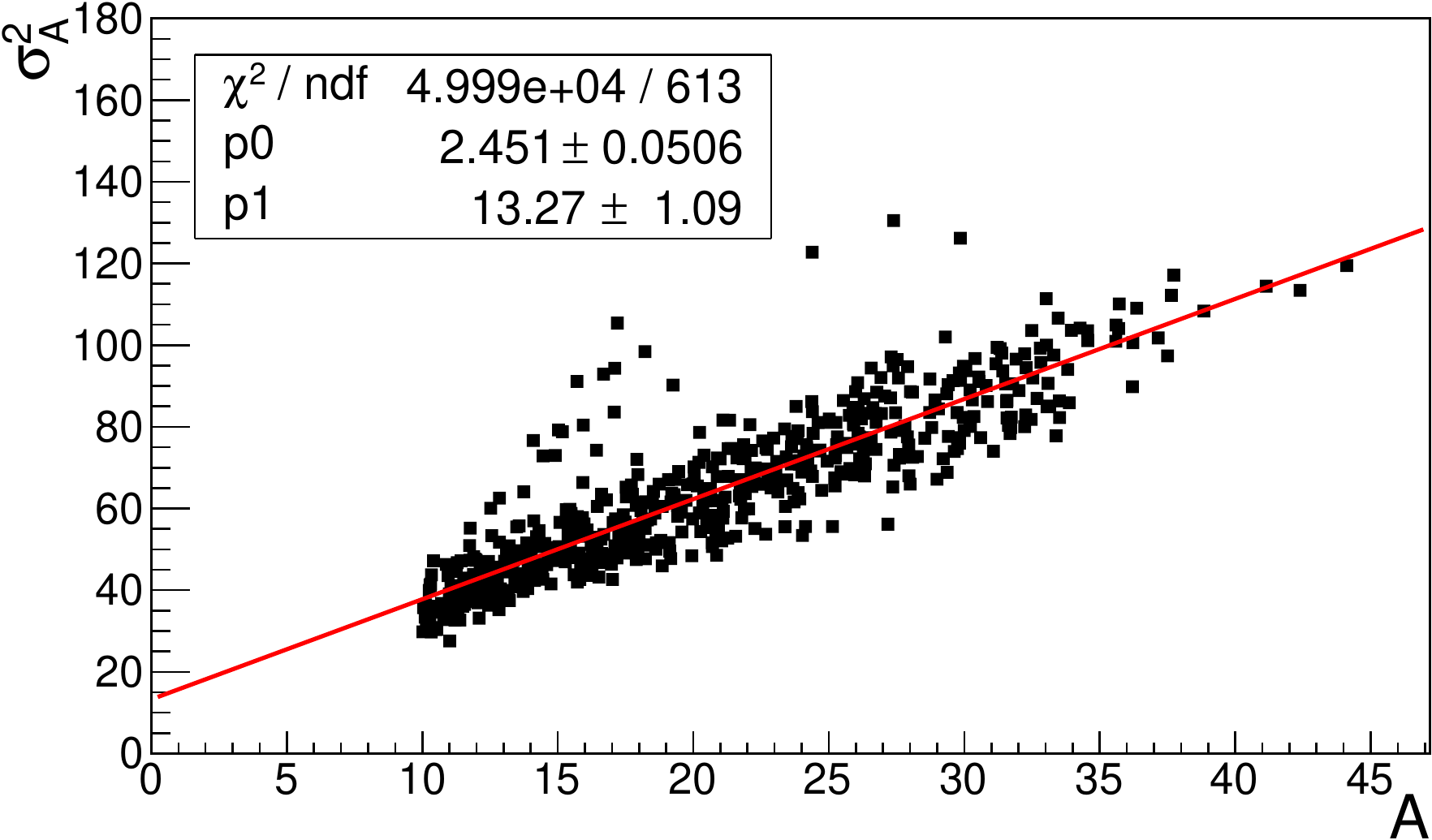}\quad\fig{0.48}{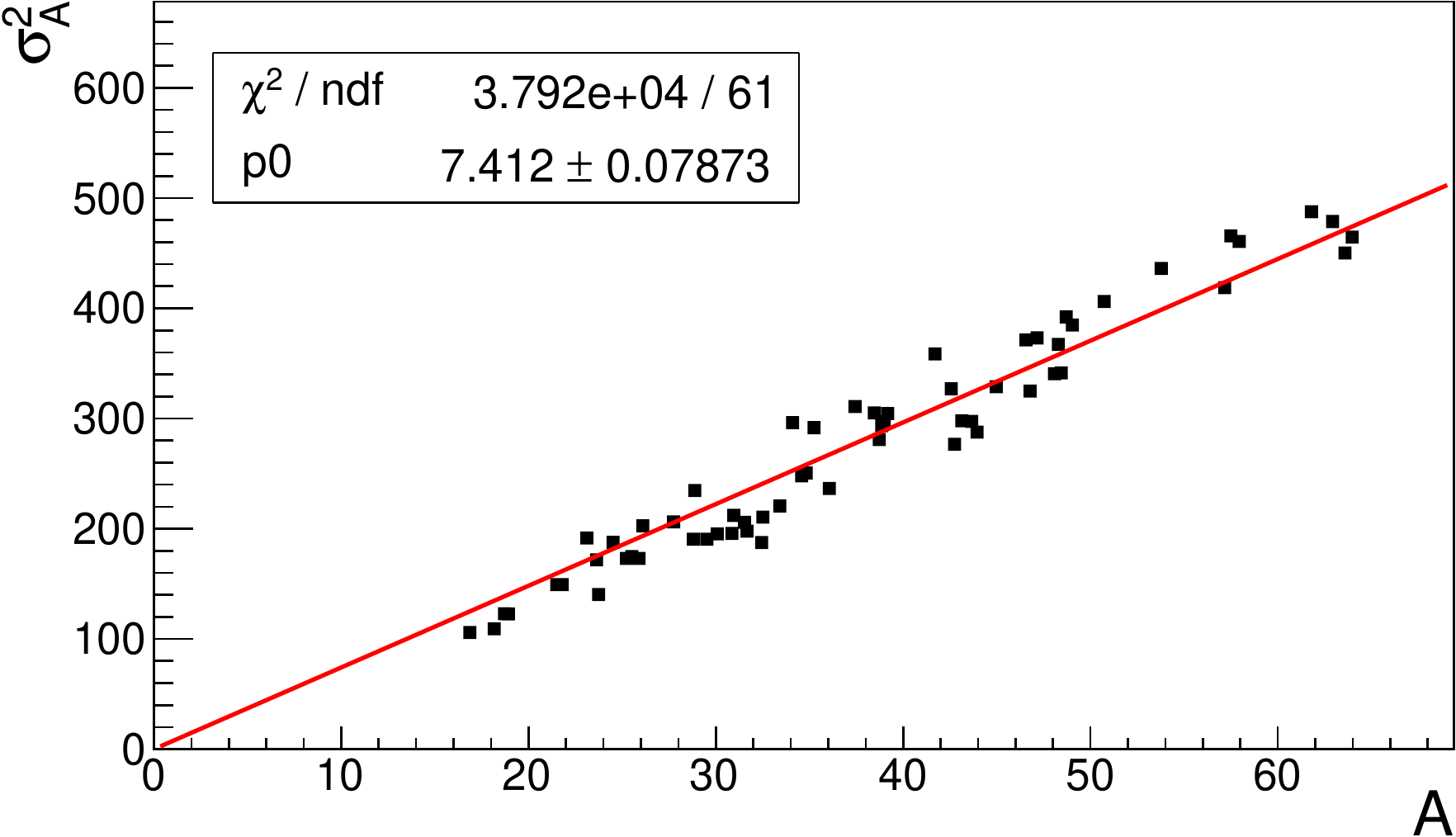}}
   \caption{Linear approximations of the $\sigma^2_A(A)$ dependence for
   channels~(11,2) and~(14,3). See the text for details.}
   \label{fig:fit161and211}
\end{figure}

To estimate the gain~$G$ of PMTs, we used the fact that the variance of
a digital signal~$\sigma^2_A$ is a linear function of its mean value~$A$
for a (quasy)stationary input signal on the photocathode.  Denoting
by~$p_0$ the slope coefficient of the $\sigma^2_A(A)$ dependence, the
product of the gain~$G$ and the anode voltage-to-ADC code coefficient~$a$
can be written as $aG=2Cp_0/q_\mathrm{e}$, where $C=30$~pF is the anode
capacitance, cf.\ Eq.~(\ref{eq:s-epsG}).
Two examples of a linear approximation of the $\sigma^2_A(A)$ dependence
are shown in Figure~\ref{fig:fit161and211}. Events with stationary
signals were selected from the TUS data for this procedure.
PMT gains for the ten hit channels of the TUS161003 event were estimated
using this method.
They are
presented in the first row of Table~\ref{tab:abssense}, see also
Figure~\ref{fig:focal_surface}.

\begin{table}[!ht]

	\caption{PMT gains, relative optical efficiencies and absolute
	sensitivities of the hit channels of the TUS161003 event. Modules and
	channels of the photodetector are denoted as (md,ch).}

	\begin{center}\small
	\begin{tabular}{|c|c|c|c|c|c|c|c|c|c|c|}
		\hline	
		(md,ch)& (11,1) & (11,2) & (12,1) & (12,2) & (12,3) & (13,1) & (13,2) & (13,3) & (13,4) & (14,3) \\
		\hline
		$G\times10^{-6}$ &  1.14 & 1.79& 1.34& 2.33& 3.86& 1.26& 2.00& 1.12& 1.23& 5.42 \\
	   \hline
   	$\epsilon$, rel.&  1.26 & 1& 0.63& 0.45& 0.51& 1.26& 0.49& 0.70& 0.71& 0.86 \\
	   \hline
		$s$, $\mu$s$\cdot$m$^2 $&  0.80 & 1.00& 0.47& 0.59& 1.12& 0.90& 0.55& 0.44& 0.49& 2.61 \\ 
	\hline
	\end{tabular}
	\end{center}
	\label{tab:abssense}
\end{table}

We employed two considerations to estimate the optical efficiency of the
hit channels: (i)~stationary signals have very small relative
fluctuations in the ``Meteor'' mode since the signal in this mode is
integrated during 8192 time steps of the main mode, and (ii)~the FOV of
a channel shifts by 13~km in 1.68~s, during which a record in the
``Meteor'' mode was obtained, so that an area observed by a channel
at the beginning of the record is later observed by an adjacent one,
next with respect to the direction of movement of the satellite.  This
way, one can select signals which are stationary in time and (partially)
uniform in the FOV and use them to estimate the relative sensitivity of
channels.  The optical efficiency~$\epsilon$ of the hit channels
relative to channel (11,2) is presented in the second row of
Table~\ref{tab:abssense}.

Finally, absolute sensitivities of all hit channels are given in the
third row of Table~\ref{tab:abssense}. Here, Eq.~(\ref{eq:s-epsG}) was used
together with an estimate of the average optical efficiency of~0.14.

We remark that the above method of in-flight calibration only
takes into account 
fluctuations in the number of photoelectrons on a photocathode.
However, the ADC code variance can be influenced by various other
factors such as fluctuations of the secondary electron emission in the
dynode system, voltage variations on the PMT high voltage power supply,
an additional noise in the analog electronics and digitization path.
These aspects as well as other details of the method and its results
will be addressed in a dedicated publication.

\section{Weather conditions of the observation}
\label{app:weather}

\begin{figure}[!t]
    \centering
	\includegraphics[width=0.9\textwidth]{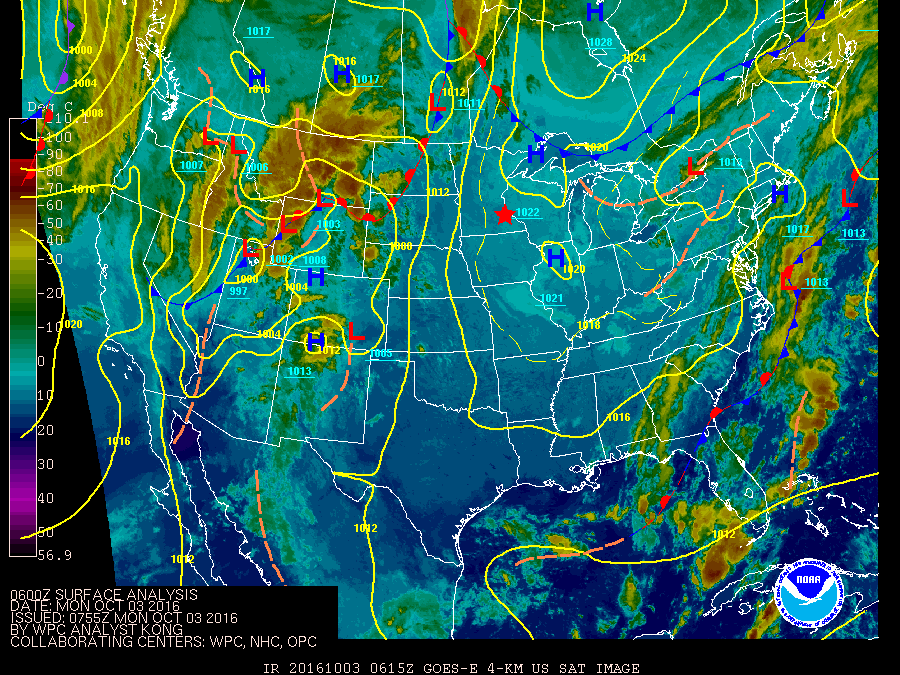}

	\caption{Satellite image of the U.S.\ with surface pressure isobars
	(yellow lines), fronts (blue triangled and red circled lines), high
	and low pressure areas and state borders (white lines). The RGB
	composite could be summarised as follows: dark blue is clear sky,
	light blue are low clouds, green and yellow are medium clouds, dark
	yellow-orange are high or vertical developing clouds.
	The star shows the location of the TUS161003 event.}

	\label{fig:sfc06}
\end{figure}

The first map considered (Fig.~\ref{fig:sfc06}) was the analysis chart
with fronts and the RGB satellite composite image at 6:00 UTC, i.e.,
in 11 minutes after the TUS161003 event.
A ridge, with its central high pressure located on the Western Canada,
was extended roughly north-south on the Eastern part of the U.S.\ and on
the Great Lakes area.  On the other side, a trough dominated the Rocky
Mountains and Western Plains area, and it was characterised by several
local pressure minima.  An extensive frontal zone passed through the
central-western states, and progressively moved towards East, pushed by
a high pressure field located on the Western Coast.

The map in Fig.~\ref{fig:sfc06} shows that
a series of low-pressure systems
were developing on the central-western U.S.\
that was organised into a thunderstorm line proceeding from West to
East and positioned above Dakota  18 hours later.
However, there were no clouds on the Eastern Minnesota and Wisconsin
at 5:49 UTC as shown by Fig.~\ref{fig:sfc06} and also by the GEOS~13
image (at 5:45 UTC not shown),
while fog was present in southern Wisconsin,  which is depicted
with light colours.

\begin{figure}[!t]
	\centering
	\fig{.55}{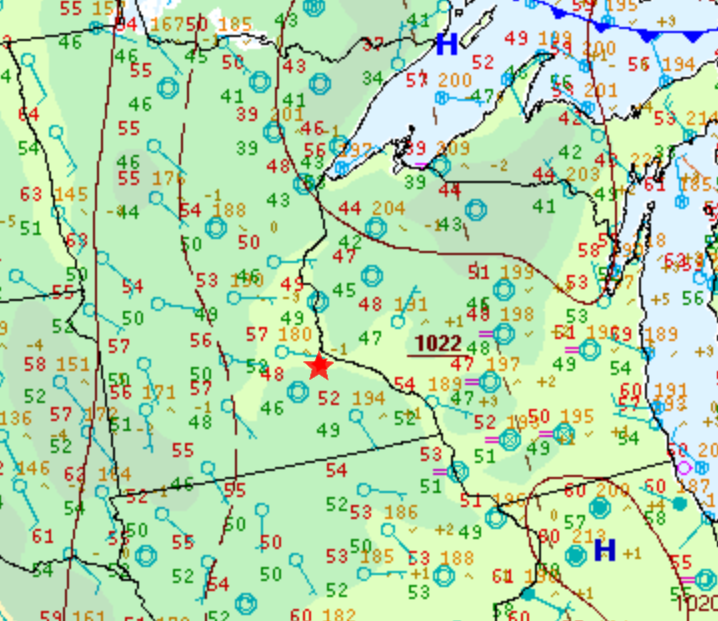}

	\caption{Synoptic weather stations between Minnesota and Wisconsin.
	The light-blue circles represent the cloud-cover in octas, and from
	the circles is drawn the wind barb, which tail indicates the wind
	speed in knots (5, 10, 50 \si{\knot}). The red number (top-left corner)
	is the 2~m temperature (degrees Fahrenheit), the green number
	(bottom-left corner) is the dew point temperature (degrees
	Fahrenheit), the yellow number in the top-right corner is the sea
	level pressure in tenths of \si{\hecto\pascal}, with the leading 10
	or 9 omitted (i.e. 1013.2 \si{\hecto\pascal} is 132). Below the circle
	is plotter the cloud type if any, on the left is the pressure trend, on
	the right are the weather phenomena at observation time.
	The star shows the location of the TUS161003 event.}

	\label{fig:stazioni}
\end{figure}

The atmosphere sounding at disposal was launched at Chanhassen
(\ang{44;51;44} N and \ang{93;31;50} W), and shows a not saturated,
practically moist adiabatic profile with different layers of inversions.
The tropopause was located at \SI{200}{\hecto\pascal} around
\SI{12200}{\metre}.  Unfortunately, the nearest sounding was at 00~UTC on
3\textsuperscript{rd} October (late afternoon) and the next one at 12:00~UTC
that corresponds to the morning local time.  The evolution of the atmosphere
profile was indeed coherent with a night thermal inversion building, and
ideally at 6:00~UTC the inversion was partially built.  This could
justify the presence of fog at some stations
(Fig.~\ref{fig:stazioni}), but also the clear sky conditions due to
radiative cooling of the ground.

The station observations of the area interesting for the event confirm
the aforementioned (Fig.~\ref{fig:stazioni}), and highlight the
presence of light fog in the southern Wisconsin, but not in the area
where the event was detected. The wind was feeble or calm,
\SI{2.6}{\metre\per\second} or less, and the average temperature ranged
between \SI{7}{\celsius} (44 degrees Fahrenheit) and \SI{14}{\celsius}
(57 degrees Fahrenheit).  Some stations in the south part of Wisconsin 
had observed light fog (at other locations the fog was not observed but
the air was saturated, in fact the dew point temperature was the same of
air temperature).  The sea level pressure trends were distributed as
follows: a drop of pressure was
generally measured in the western part of Minnesota,
while there was a general increasing or stability on Wisconsin.

\begin{figure}[!t]
	\centering
	\fig{.7}{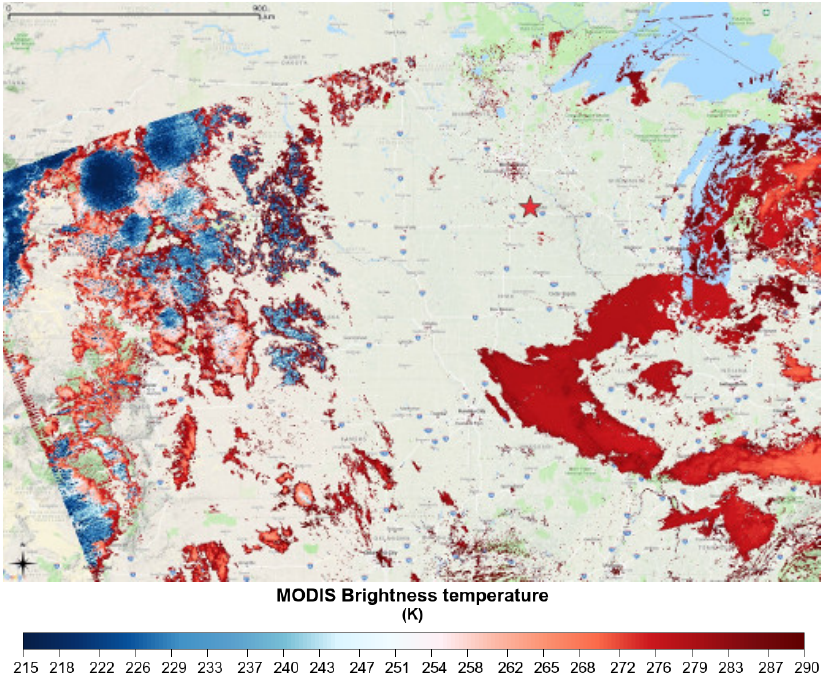}

	\caption{MODIS cloud fraction at 4:20 UTC obtained with a specific
	algorithm. One can see a Croatia-shape fog cloud
	centered at \ang{40}~N \ang{92}~E.
	The star shows the location of the TUS161003 event.}

	\label{fig:modis}
\end{figure}

\newpage
Looking at the MODIS satellite, that had shot an image at 4:20 UTC
(Fig.~\ref{fig:modis}), one can see the fog (shown in red) but also
other clouds that probably were wrongly detected or that were not present
at 6:00 UTC when operators compiled the SYNOP messages
(Fig.~\ref{fig:stazioni}).

The cold front is not already formed, and the low pressure system is far
from the event site.  The Chanhassen station mentioned above showed that
a strong inversion built up during the night, and this confirms the
hypothesis of a clear sky. In fact, the ground could irradiate a lot of
energy, and it reached low temperature values.


\providecommand{\href}[2]{#2}\begingroup\raggedright\endgroup

\end{document}